\documentclass[aps,preprint,superscriptaddress,floatfix,12pt,nofootinbib]{revtex4-2}
\usepackage[utf8]{inputenc}
\usepackage{amsmath,amssymb}
\usepackage{empheq}
\usepackage{mathtools,cuted}
\usepackage{multirow}
\usepackage{xcolor}
\usepackage[export]{adjustbox}
\usepackage{graphicx}
\usepackage{dcolumn}
\usepackage{bm}
\usepackage{hyperref}
\usepackage{physics}
\usepackage[mathlines]{lineno}
\usepackage[capitalise]{cleveref}
\usepackage{multirow}
\usepackage{bbm}
\usepackage{orcidlink}
\usepackage[inkscapelatex=false]{svg}
\usepackage{float}
\usepackage{braket}

\begin{document}

\title{Ubiquitous topological materials protected by hidden partial symmetries}

\author{Lumen Eek\,\orcidlink{0009-0009-1233-4378}}
\affiliation{Institute of Theoretical Physics, Utrecht University, Utrecht, 3584 CC, Netherlands}

\author{Xuelong Chen\,\orcidlink{0009-0002-3885-5402}}{}
\affiliation{Eastern Institute for Advanced Study, Eastern Institute of Technology, Ningbo, China}
\affiliation{Department of Electrical and Electronic Engineering, The Hong Kong Polytechnic University, Hong Kong, China}

\author{Malte Röntgen\,\orcidlink{0000-0001-7784-8104}}\email{mroentgen@eitech.edu.cn}
\affiliation{Eastern Institute for Advanced Study, Eastern Institute of Technology, Ningbo, China}

\author{Wenlong Gao\,\orcidlink{0000-0002-3446-7157}}\email{wgao@eitech.edu.cn}
\affiliation{Eastern Institute for Advanced Study, Eastern Institute of Technology, Ningbo, China}

\author{Cristiane Morais Smith\,\orcidlink{0000-0002-4190-3893}}
\affiliation{Institute of Theoretical Physics, Utrecht University, Utrecht, 3584 CC, Netherlands}

\date{\today}

\begin{abstract}
\textbf{Symmetries play a crucial role in protecting the topological phases of insulators, ensuring the robustness of their surface states against perturbations.
Traditionally, these symmetries have to be obeyed by the system as a whole, which is a rather strong constraint.
In this work, we demonstrate that topological effects can also be realized through a substantially weaker form of symmetry that we coin \emph{partial latent symmetries}.
For these, invariance is required only at certain energy scales of an effective Hamiltonian, rather than for the object as a whole.
Applying this idea, we construct weak-symmetry versions of the Su-Schrieffer-Heeger and Benalcazar-Bernevig-Hughes models.
These weak-symmetry models behave in many aspects just like their conventional brethren---in particular, they host robust edge states---, though the corresponding topological invariants are only \emph{approximately} quantized.
We conclude this work by experimentally realizing Hermitian and non-Hermitian versions of such approximate topological insulators and probe their edge states.
Overall, given the ubiquity of symmetry in all areas of physics and engineering, we foresee our work to serve as a cornerstone for exploring how conventional, strong symmetry requirements can be replaced by weak-symmetry ones.
}
\end{abstract}
\maketitle{}

\section*{Introduction}

\begin{figure}[htb]  
\centering
\includegraphics[width=0.5\columnwidth]{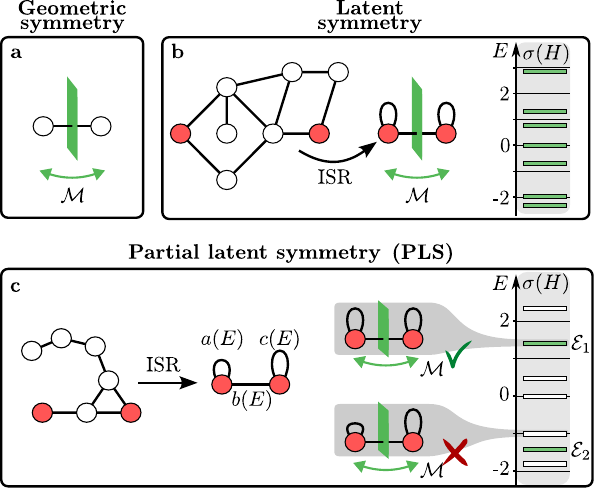} 
\caption{
\textbf{Progression of symmetry concepts.} 
\textbf{a}, Conventional mirror symmetry characterized by the presence of a mirror plane. 
\textbf{b}, Latent symmetry: a system is said to be latent symmetric if its ISR possesses a symmetry at all energies, here exemplified by mirror symmetry. 
\textbf{c}, Partial latent symmetry: upon performing an ISR, the system exhibits a symmetry only for a subset of its eigenstates, i.e., the system is symmetric only at specific energies (marked in green), here at $\mathcal{E}_{1,2}=\pm\sqrt{2}$.
Throughout this figure, each line in the original, unreduced system is equal to unity.
Expressions for the energy-dependent terms appearing in the ISR are given in the Methods section.
}
\label{fig:overviewFigure}
\end{figure}
From Lorentz invariance \cite{Peskin2018IntroductionQuantumFieldTheory} to Emmy Noether's explanation of conservation laws \cite{Noether1918NVGWZGMK1918235InvarianteVariationsprobleme}, symmetries have long provided the scaffolding on which contemporary physics is built.
Modern condensed-matter physics, for instance, heavily relies on Bloch's famous theorem of 1929 that links translational symmetries to band structure \cite{Bloch1929ZP52555UeberQuantenmechanikElektronenKristallgittern}.

In recent decades, the notion of symmetry has also become central to the classification of topological materials. 
These are insulating in the bulk, but metallic or superconducting at the edges \cite{Bernevig2013TopologicalInsulatorsTopologicalSuperconductors}. The metallic edge states are dissipationless as long as they are protected by the underlying preserved symmetries.
A major breakthrough is the tenfold classification, which,---based on time-reversal, charge conjugation, and chiral symmetry---successfully predicts the existence and type of topological states purely based on the dimensionality and symmetries of non-interacting systems \cite{altlandNonstandardSymmetryClasses1997, kitaevPeriodicTableTopological2009}.
Later, this tenfold way was further enriched by adding other symmetries; particularly, spatial symmetries such as inversion or rotations.
This lead to the discovery of higher-order topological insulators (HOTIs), which rely on spatial symmetries to host corner states (for 2D and 3D materials) \cite{benalcazarQuantizedElectricMultipole2017,song$densuremath2$DimensionalEdgeStates2017, langbehnReflectionSymmetricSecondOrderTopological2017, Benalcazar2019PRB99245151QuantizationFractionalCornerCharge}, and hinge states (for 3D materials) \cite{schindlerHigherorderTopologicalInsulators2018}.
As powerful as this extended classification is, many systems are devoid of the symmetries that would offer topological protection. 
From a practical perspective, given the large number of such asymmetric systems, the situation is rather unsatisfying.

Recently, an entirely new perspective on seemingly asymmetric systems has been provided through the framework of \emph{latent symmetries} \cite{Smith2019PA514855HiddenSymmetriesRealTheoretical}.
A latent symmetry is not visible in the original system, but can be uncovered by performing an isospectral reduction (ISR) \cite{Bunimovich2014IsospectralTransformationsNewApproach}.
Originally proposed in graph theory, latent symmetries have recently been used in physics as well \cite{Rontgen2023PRL130077201HiddenSymmetriesAcousticWave,Rontgen2021PRL126180601LatentSymmetryInducedDegeneracies,Cui2023PRL131237201ExperimentalRealizationStableExceptional,himmelStateTransferLatentsymmetric2026}.
In the context of topology, latent symmetries have been shown to be completely equivalent to spatial symmetries \cite{Bunimovich2014IsospectralTransformationsNewApproach,eekHigherorderTopologyProtected2025,moustajLatentHaldaneModels2025,zhangLatentGeometryCorrespondenceUnraveling2026}.
In particular, both corner and hinge states can be successfully realized in systems protected by latent symmetries. 

In this work, we fuse the concept of latent symmetry with the idea of symmetries being present only at certain energy scales.
The result is a generally applicable framework of what we call \emph{partial latent symmetry}, which drastically enlarges the group of systems that should be considered symmetric. 
We further link this framework to topology by showing that it leads to unexpected robust topological behavior, thereby breaking several paradigms and showing that much less is required to generate topological states than it is currently believed.

\begin{figure*}[t]  
\centering
\includegraphics[width=\textwidth]{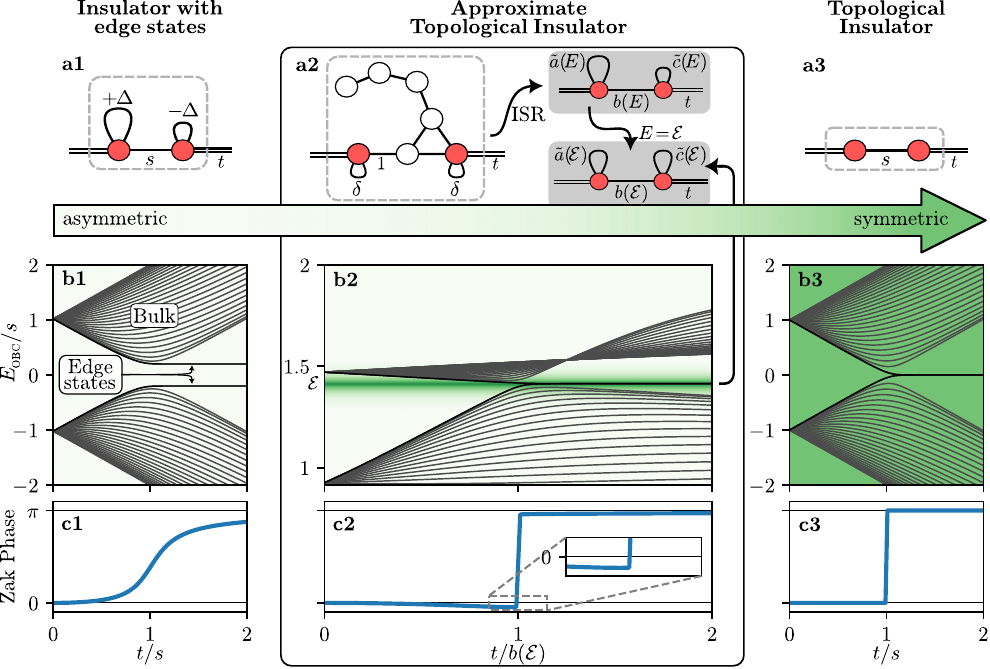} 
\caption{
\textbf{Edge states in one-dimensional (topological) insulators.} 
\textbf{a1--a3}, Schematic unit cells. \textbf{a1} Sublattice-imbalanced chain with alternating hopping amplitudes. 
\textbf{a2} One-dimensional chain exhibiting a \emph{partial latent symmetry}. Upon performing the ISR, an energy-dependent sublattice-imbalanced effective chain is obtained. At the energy $E=\mathcal{E}$, the imbalance vanishes and the system exhibits a partial latent mirror symmetry. 
\textbf{a3} Su-Schrieffer-Heeger chain with a mirror-symmetric unit cell. 
\textbf{b1--b3}, Energy spectra as a function of $t/s$ for the corresponding chains with open boundary conditions. The background color indicates the degree of sublattice imbalance at a given energy. For the models in \textbf{a1} and \textbf{a3}, the imbalance is energy independent, whereas for the partial latent symmetric system it is energy dependent. The arrow above the panels serves as a color bar, with white indicating strong imbalance and green indicating vanishing imbalance, corresponding to mirror symmetry. 
\textbf{c1--c3}, Zak phase of the occupied bulk bands as a function of $t/s$. \textbf{c1} In the absence of mirror symmetry, the Zak phase varies continuously and is not quantized. 
\textbf{c2} In the presence of a partial latent symmetry, the Zak phase is approximately quantized and exhibits an abrupt change at the gap closing; deviations from perfect quantization are highlighted in the inset. 
\textbf{c3} With exact mirror symmetry, the Zak phase is quantized to $0$ or $\pi$, consistent with the conventional bulk--boundary correspondence.
}
\label{fig:fig2}
\end{figure*}

\section*{Evolution of symmetry concepts}
Most topological materials rely on symmetry to protect their topological phase. 
A particularly simple example is the Su-Schrieffer-Heeger (SSH) model \cite{Su1979PRL421698SolitonsPolyacetylene}.
Its unit cell, depicted in \cref{fig:overviewFigure}a, is mirror symmetric, that is, the corresponding Hamiltonian $H$ commutes with the symmetry operation $\mathcal{M}$,
\begin{equation}
\mathcal{M} H \mathcal{M}^{-1} = H \,.
\label{eq:sym}
\end{equation}
This mirror symmetry, or, alternatively, the system's chiral symmetry, leads to a quantized Zak phase \cite{Zak1989PRL622747BerrysPhaseEnergyBands}, which, for sufficiently strong inter-cell hopping, acquires a non-trivial value.
As a consequence, mirror symmetry protected topological edge states emerge.

Generally speaking, any one-dimensional chain with a mirror-symmetric unit cell has a quantized Zak phase, thereby providing a convenient mechanism for the design of topological systems.
As powerful as the connection between symmetry and quantized topological invariants is, demanding a setup to be symmetric is a rather strong constraint.
Recently, it has been shown that topological insulators may also be protected by a weaker and hidden form of symmetry, a latent symmetry \cite{Rontgen2024PRB110035106TopologicalStatesProtectedHidden,eekHigherorderTopologyProtected2025}.
A latent symmetry becomes visible after performing the ISR $\mathcal{R}_S[H,E] = H_{SS} + H_{S\overline{S}} (E \mathbb{I}-H_{\overline{SS}})^{-1} H_{\overline{S}S}$, where $H$ is the Hamiltonian partitioned into the two sets of sites $S$ and its complement $\overline{S}$.
Mathematically speaking, latent symmetry can be expressed as
\begin{equation}
\mathcal{M}\mathcal{R}_{S}[H,E]\mathcal{M}^{-1} = \mathcal{R}_{S}[H,E]\,. \label{eq:lat_topo}
\end{equation}
A simple latent symmetric model, along with its ISR, is depicted in \cref{fig:overviewFigure}b.

In the following, we dilute the constraints that are put on topological materials even further.
Since a significant portion of the interest in topological materials arises from their symmetry protected edge states, our key idea is to \emph{engineer systems that exhibit symmetry solely at the energy corresponding to the edge states, while not requiring symmetry elsewhere}.

Formalizing this idea, we introduce the notion of a \textit{partial latent symmetry}. Just as for a latent symmetry, it concerns a symmetry relation on the level of the ISR. However, unlike Eq.~\eqref{eq:lat_topo}, a partial latent symmetry does not hold for arbitrary values of $E$. Instead, it only holds for one or more eigenvalues $\mathcal{E}$ of $H$. That is,
\begin{equation}
\mathcal{M}\mathcal{R}_{S}[H,\mathcal{E}]\mathcal{M}^{-1} = \mathcal{R}_{S}[H,\mathcal{E}]\,. \label{eq:partial_sym}
\end{equation}
In \cref{fig:overviewFigure}c, we showcase a simple seven-site Hamiltonian that has two partial latent symmetries at energies $\mathcal{E}_{1,2} = \pm \sqrt{2}$.
At these two energies, the ISR becomes
\begin{equation}
    \mathcal{R}_{S}[H,\mathcal{E}_{1,2}] =\begin{pmatrix}
        a(\mathcal{E}_{1,2}) & b(\mathcal{E}_{1,2}) \\
        b(\mathcal{E}_{1,2}) & c(\mathcal{E}_{1,2})
    \end{pmatrix} =  \pm \frac{1}{\sqrt{2}}\begin{pmatrix}
        1 & 1 \\
        1 & 1
    \end{pmatrix} \,.
\end{equation}

\Cref{eq:partial_sym} is a generalization of the similar concepts of partial symmetry \cite{Gozdz2012PAN751195SpectralTheoremPartialSymmetries,Pedrak2014PS89054024IntrinsicHamiltonianSymmetryGroup}, quasi symmetry \cite{Ren2021PRL126120604QuasisymmetryGroupsManyBodyScar}, and emergent symmetry \cite{batistaAlgebraicApproachInteracting2004a,nussinovSymmetryPrincipleTopological2009}.
It is further a generalization of latent symmetries, for which it would hold for all eigenvalues, and even of conventional symmetries, for which the ISR would be taken over all sites, and then naturally becomes the original Hamiltonian $H$.

Determining how many real materials feature partial latent symmetries is an interesting question for future research.
This endeavour would require considerable effort such as high-throughput calculations, akin to those done for topological materials \cite{vergnioryAllTopologicalBands2022,wiederTopologicalMaterialsDiscovery2022}.
For this work, we resorted to a reasonable surrogate and analysed large numbers of graphs instead.
To this end, we generated approximately 12 million different graphs, each of them equivalent to a tight-binding Hamiltonian (unit cell) of dimension up to $10$.
Overall, the majority of these graphs (60 percent) feature a partial latent symmetry, while only 33.5 percent of them are geometrically symmetric.
If these unexpected results hold for real materials, even if only approximately, it means that our usual classification into symmetric and asymmetric systems is far from complete and must be reconsidered.

\section*{One-dimensional approximate topological insulators}
In the following, we apply our new symmetry concept to the realm of topology.
To set the stage, we first review the two extreme examples of total asymmetry and total symmetry; partial latent symmetry will fit right between those. In \cref{fig:fig2}a1, we show a bipartite chain with alternating hopping $s$ (intra-cell) and $t$ (inter-cell),  and imbalanced sublattice potential $\pm \Delta$; this system is totally asymmetric.
For sufficiently strong inter-cell hopping $t$, a finite realization of this chain hosts exponentially localized edge states (red) with energies $E\approx \pm \Delta = \pm 0.2 s$, see \cref{fig:fig2}b1.
These edge states persist over a wide parameter range, but are shifted away from mid-gap and are therefore not symmetry-protected in the usual sense.
This lack of pinning can be traced back to the absence of a quantized bulk topological invariant.
In particular, the Zak phase of this model varies continuously as a function of $t$ and is not restricted to the two values $0$ and $\pi$, see \cref{fig:fig2}c1.

The situation changes qualitatively once mirror symmetry of the unit cell is restored, i.e., by eliminating the sublattice imbalance $\Delta$.
The corresponding system is depicted in \cref{fig:fig2}a3; it is the SSH model discussed above.
Importantly, the Zak phase is now quantized [\cref{fig:fig2}c3].
When it jumps from $0$ to $\pi$ at $s=t$, where the inter-cell hopping $t$ becomes equal to the intra-cell hopping $s$, edge states emerge, as shown in \cref{fig:fig2}b3.
Contrary to the asymmetric case, these states are now (approximately) degenerate and pinned at the same energy.
This is a consequence of the classical bulk–boundary correspondence: A quantized Zak phase in the bulk enforces pinned, symmetry-protected boundary modes.
We remark that the bulk-boundary correspondence would also hold for a unit cell that is not mirror-symmetric, but latently mirror-symmetric.

We now use our new symmetry concept to fill the gap between symmetry and asymmetry.
To this end, we use the partial latent symmetric model of \cref{fig:overviewFigure}c as a unit cell, cf. \cref{fig:fig2}a2.
For reasons that become clear in a moment, we equip the two sites on which the ISR is performed with an additional on-site potential $\delta = 1/\sqrt{2}$.
This shifts the effective on-site potentials as $\tilde{a}(E) = a(e) + \delta$, $\tilde{c}(E) = a(e) + \delta$  and leaves $b(E)$ invariant.
In particular, this change preserves partial latent symmetry.
Upon performing the ISR, one obtains a sublattice-imbalanced effective model similar to \cref{fig:fig2}a1.
Now, however, the imbalance $\tilde{a}(E) - \tilde{c}(E)$ is energy-dependent and, due to partial latent symmetry, vanishes completely at $E=\mathcal{E}_{1,2}$.
In the following, we focus on $\mathcal{E}_1 = \sqrt{2} := \mathcal{E}$.

With our choice of $\delta$, we additionally have $\tilde{a}(\mathcal{E}) = \tilde{c}(\mathcal{E}) = \mathcal{E}$.
At energy $\mathcal{E}$, the system is thus an effective SSH model with identical on-site potential $\mathcal{E}$ and hopping $b(\mathcal{E})$.
By analogy to the SSH model, it can be shown that edge states emerge at this very energy $E = \mathcal{E}$, provided that the inter-cell hopping $t$ is larger than the effective intra-cell hopping $b(E=\mathcal{E})$.
Importantly, both edge states are now pinned at the same energy $\mathcal{E}$; cf. \cref{fig:fig2}b2.

Interestingly, the Zak phase of this system is only \textit{approximately quantized}, see \cref{fig:fig2}c2. This reflects the fact that the Zak phase is a bulk property, computed from extended Bloch states. While the edge states themselves reside in the partial latent symmetric subspace, the bulk states do not. As a consequence, the Zak phase, evaluated from bulk bands, need not be strictly quantized. Nevertheless, it remains close to $\pi$ throughout the topological regime and changes abruptly at the gap closing, marking a topological phase transition. 

From a bulk perspective, this behaviour can be understood as follows: Exact quantization of the Zak phase requires a symmetry that constrains the Bloch Hamiltonian at all energies and crystal momenta. In the present case, such a constraint is only satisfied at the isolated energy $E=\mathcal{E}$, where the partial latent symmetry emerges.
The bulk bands, therefore, only approximately respect the symmetry, leading to an approximately, but not exactly, quantized Zak phase.
If one instead evaluates the ISR at $E=\mathcal{E}$ and calculates the Zak phase for the resulting effective Hamiltonian, one obtains a fully quantized Zak phases (see Sec. S1C of the Supplementary Information).

\section*{Experimental realization of the approximate BBH model}

\begin{figure*}[t]  
\centering
\includegraphics[width=0.9\textwidth]{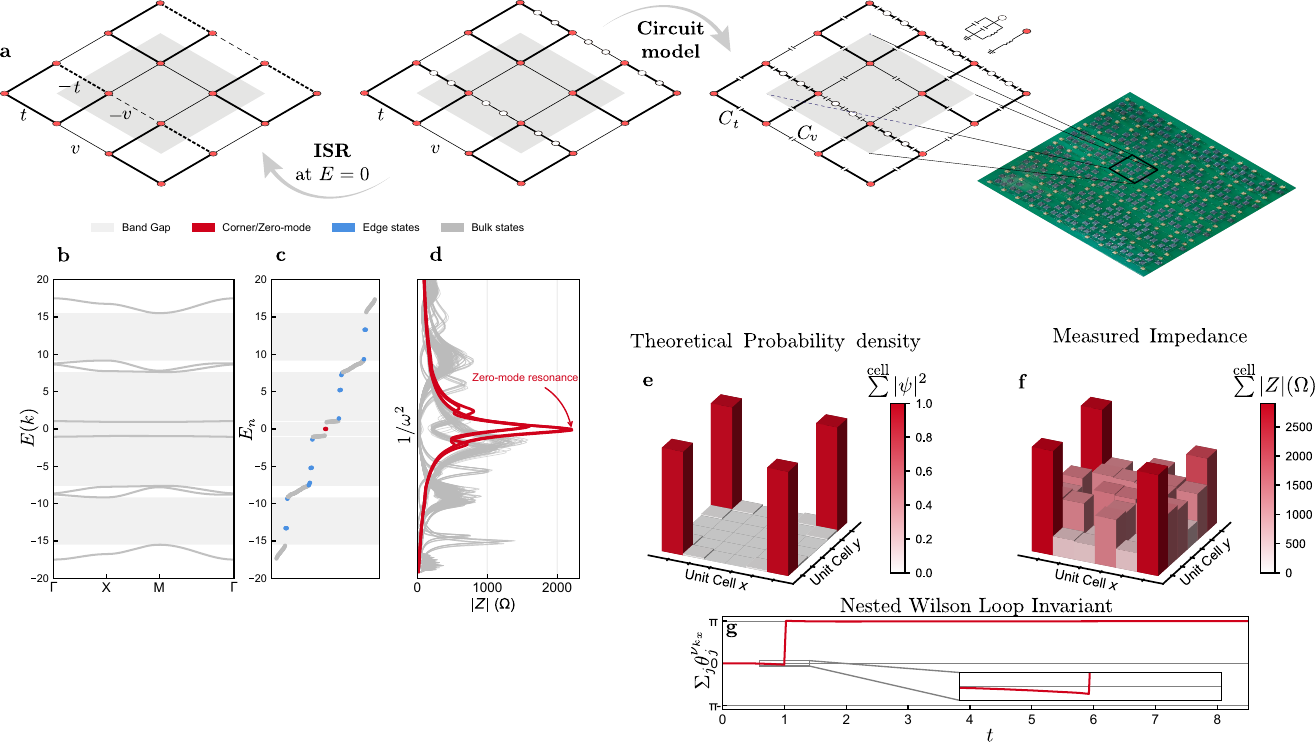} 
\caption{\textbf{Topological circuit realization and characterization.}
\textbf{a}, Schematic illustration of the topological circuit realizing a two-dimensional corner state tight-binding model.
The left structure shows the ISR of the tight-binding lattice model (second structure) at $E=0$, while the right structure depicts the physical circuit board implementing the model. An intermediate structure illustrates the mapping from the tight-binding links to the circuit components.  
\textbf{b}, Projected band structure ($E$ vs.\ $k$) of the system with periodic boundary conditions. The parameters used are the intra-cell hopping $v = 1$ and inter-cell hopping $t = 8.2$. The gray areas denote the bulk states, and the white areas are the band gaps.  
\textbf{c}, Energy spectrum ($E_n$) of the $6 \times 6$ unit cell system (276 sites) with open boundary conditions. The bulk states are in gray, edge states in blue, and the corner/zero-modes are highlighted in red.  
\textbf{d}, Experimental measurement of the inverse squared frequency ($1/\omega^2$) versus the measured impedance $|Z|$ ($\Omega$). The parameters for the topological circuit are an intra-cell capacitance of 100~pF and an inter-cell capacitance of 820~pF. The red peak marks the zero-mode resonance.  
\textbf{e}, Calculated spatial distribution of the unit-cell-resolved probability density, defined as $\rho(x_\text{cell}) = \sum_{i \in x_\text{cell}} |\Psi_i|^2$. This summation is performed over the internal degrees of freedom within each of the 36 unit cells for the four corner states ($n=137$ to $140$), revealing strong corner localization.  
\textbf{f}, Experimentally measured spatial profile of the summed impedance within each unit cell at the resonance frequency ($f \approx 1.259$~MHz). The concentration of the integrated response at the lattice corners confirms the emergence of the predicted topological corner phase. \textbf{g}, Nested Wilson Loop Invariant for the partial latent symmetric BBH model.}
\label{fig:fig3}
\end{figure*}

After this introductory example, let us now extend our discussion to a more complicated case, namely, a two-dimensional realization of an approximate higher-order topological insulator supporting corner-localized states. The second structure in \cref{fig:fig3}a shows a tight-binding model with two different hoppings $t$ and $v$, and with the shaded region indicating a single unit cell.
At $E=0$---which, in the thermodynamic limit, is an eigenvalue of this setup---, the
system becomes exactly equal to the classic Benalcazar-Bernevig-Hughes (BBH)-type model \cite{benalcazarQuantizedElectricMultipole2017}.
Our system thus has a partial latent symmetry identical to the symmetry of the original BBH model.
In contrast to the latter, which requires negative hopping amplitudes corresponding to an effective $\pi$ flux, however, the hopping parameters in our system are \textit{real} and \textit{positive}.

We experimentally realized our model using electric circuits, where the BBH model has previously been realized as well \cite{Imhof2018NP14925TopolectricalcircuitRealizationTopologicalCorner}.
In the right part of \cref{fig:fig3}a, we show the layout and circuit-board implementation of a $6\times6$ unit cell lattice with open boundary conditions.
In the experiment, $t$ and $v$ hoppings are implemented using capacitances of $100$~pF and $820$~pF, respectively.
Additional details of the circuit realization are provided in the Methods.
Figs.~\ref{fig:fig3}~b,c display the band structures for periodic and open boundary conditions, respectively, for $v = 1$ and $t=8.2$.
The bulk bandgaps are indicated by shaded regions. Under open boundary conditions, in-gap states appear: edge states and corner states are distinguished in blue and red, respectively.
In the central gap, a quartet of nearly degenerate states emerges, corresponding to corner-localized modes of the finite system.
Experimentally, these states manifest as a pronounced zero-mode resonance in the impedance spectrum, shown in \cref{fig:fig3}d.
The spatial structure of these modes is illustrated in \cref{fig:fig3}e, which displays the unit-cell–summed probability density obtained from tight-binding calculations at zero energy, revealing strong localization at the four corners.
\Cref{fig:fig3}f shows the experimentally measured, unit-cell–summed impedance at the resonance frequency ($f\approx1.259$~MHz), again demonstrating localization near the lattice corners.
In contrast to the idealized calculation, the experimental maps exhibit a small but noticeable bulk contribution. This arises from finite linewidths in the measured impedance spectrum, which allow nearby bulk modes to contribute at the resonance frequency, as also visible in \cref{fig:fig3}d. Minor asymmetries in the observed corner amplitudes are attributed to unavoidable fabrication tolerances and component variations in the circuit implementation.
The higher-order topological character of the system is further supported by calculation of the nested Wilson-loop invariant, shown in \cref{fig:fig3}g (see Methods for details of calculations).
The invariant changes abruptly from near-zero to approximately $\pi$ at $v=t=1$, consistent with a transition into a corner-state phase. Because the present realization lacks the exact symmetries that protect the ideal BBH model, the invariant is not strictly quantized, similar to the behaviour observed in \cref{fig:fig2}c. Nevertheless, a partial latent symmetry at zero energy stabilizes the corner modes, allowing the higher-order topology to persist in this implementation.

\section*{Outlook}
The applicability of the theory presented here is not restricted to Hermitian topological models.
As a particular example of its generality, we show in Sec. III of the Supplementary Information the theoretical and experimental results of a non-Hermitian BBH-type model protected by partial latent symmetry.
Consequently, we foresee many interesting further applications.
This includes strong topological insulators \cite{Hasan2010RMP823045ColloquiumTopologicalInsulators}, nodal semimetals \cite{armitageWeylDiracSemimetals2018}, or, more generally speaking, degeneracies and exceptional points \cite{Heiss2012JPAMT45444016PhysicsExceptionalPoints,miriExceptionalPointsOptics2019,Meng2024APL124060502ExceptionalPointsNonHermitianSystems} protected and stabilized by partial latent symmetries.
Another venue would be topology in non-linear media, where the symmetry could be present only around a specific linearization point.

Some immediate goals for the near future would be the realization of approximate topological insulators in other systems, such as photonic waveguides \cite{Szameit2012DiscreteOpticsFemtosecondLaser} or electronic quantum simulators \cite{Weimer2010RydbergSimulator}, and the above-mentioned high-throughput search for materials hosting partial latent symmetries.

Although here we applied the concept of partial latent symmetry to topological insulators, the scope of the idea underlying this work is much more general.
In layman's terms, the aim is to \emph{relax the constraint of symmetry as far as possible while keeping the desired (material) properties}.
In this respect, our work is part of a recent trend of symmetry extensions, where symmetry holds only for some eigenstates \cite{Gozdz2012PAN751195SpectralTheoremPartialSymmetries,Ren2021PRL126120604QuasisymmetryGroupsManyBodyScar}, particularly the low-energy ones \cite{batistaAlgebraicApproachInteracting2004a}, a subset of sites \cite{Kalozoumis2014PRL113050403InvariantsBrokenDiscreteSymmetries,Wang2023NP19992SubsymmetryprotectedTopologicalStates,Rontgen2019PRB99214201LocalSymmetryTheoryResonator, rhimTopologicalProtectionLocal2026}, on average \cite{maAverageSymmetryProtectedTopological2023,maTopologicalPhasesAverage2025}, or only in lower orders of perturbation theory \cite{guoQuasisymmetryprotectedTopologySemimetal2022,huHierarchyQuasisymmetriesDegeneracies2023,liGroupTheoryQuasisymmetry2024}, to name just a few.
The time seems ripe to merge these concepts into a new, more general theory of symmetry.

\begin{acknowledgments}
M.R. acknowledges fruitful discussions with G.E. Sommer. LE and CMS acknowledge the research program “Materials for the Quantum Age” (QuMat) for financial support. This program (registration number 024.005.006) is part of the Gravitation program financed by the Dutch Ministry of Education, Culture and Science (OCW).
\end{acknowledgments}

\newpage
\section*{Methods}
\subsection*{The isospectral reduction}
To analyze effective low-energy descriptions of one-dimensional tight-binding models with complex unit cells, we employ the ISR formalism \cite{Rontgen2024PRB110035106TopologicalStatesProtectedHidden}. We partition the Hilbert space into a subset of sites $S$ and its complement $\overline{S}$, and write the eigenvalue equation $H \vec{\Psi} = E \vec{\Psi}$ in block form as
\begin{equation*}
    \begin{pmatrix}
        H_{SS} & H_{S\overline{S}}\\
        H_{\overline{S}S} & H_{\overline{SS}}
    \end{pmatrix}
    \begin{pmatrix}
        \Psi_{S} \\
        \Psi_{\overline{S}}
    \end{pmatrix} = E     \begin{pmatrix}
        \Psi_{S} \\
        \Psi_{\overline{S}}
    \end{pmatrix} \,.
\end{equation*}
Eliminating $\Psi_{\overline{S}}$, that is, the $\overline{S}$ components of $\vec{\Psi}$, we obtain the non-linear eigenvalue problem
\begin{equation} \label{eq:nlEVP}
    \mathcal{R}_S[H,E] \Psi_S = E \Psi_S\,,
\end{equation}
with
\begin{equation}
    \mathcal{R}_S[H,E]\equiv H_{SS} + H_{S\overline{S}} \left(E-H_{\overline{SS}} \right)^{-1} H_{\overline{S}S}
\end{equation}
the ISR of $H$ over $S$.
We remark that, under very mild conditions, the spectrum of $\mathcal{R}_S[H,E]$ coincides with that of the full Hamiltonian $H$ \cite{Bunimovich2014IsospectralTransformationsNewApproach}.

For the latent symmetric setup of \cref{fig:overviewFigure}b, both effective on-site potentials are equal to $a(E)$, and we have
\begin{align}
    a(E) &= \frac{2 E  ((E -1) E -1)}{((E -2) E -1) \left(E ^2+E -1\right)} \\
    b(E) &= \frac{2 E ^2}{E ^4-E ^3-4 E ^2+E +1} \,.
\end{align}

For the partial latent symmetric setup of \cref{fig:overviewFigure}c, we have
\begin{align} \label{eq:partLatSymMatElements}
    a(E) &= \frac{E ^4-3 E ^2+1}{E ^5-4 E ^3+3 E } \\
    b(E) &= \frac{E ^3-3 E +1}{(E -1) E  \left(E ^2-3\right)} \\
    c(E) &= \frac{2 E ^3-5 E +1}{(E -1) E  \left(E ^2-3\right)} \,.
\end{align}

At the two partial latent symmetry points $E=\mathcal{E}_{1,2}=\pm \sqrt{2}$, we thus have
\begin{equation}
    a = b = c = \pm 1/\sqrt{2} \,.
\end{equation}

\subsection*{Effective SSH model}
After performing the ISR, the Bloch Hamiltonian of \cref{fig:fig2}a2 reads
\begin{equation}
\mathcal{R}_S[H(k),E] =
\begin{pmatrix}
\tilde{a}(E) & b(E) + t e^{-ik} \\
b(E) + t e^{ik} & \tilde{c}(E)
\end{pmatrix},
\end{equation}
where
\begin{align}
    \tilde{a}(E) &= \frac{E ^4-3 E ^2+1}{E ^5-4 E ^3+3 E } + \delta, \\
    \tilde{c}(E) &= \frac{2 E ^3-5 E +1}{(E -1) E  \left(E ^2-3\right)} + \delta \,,
\end{align}
$b(E)$ as in \cref{eq:partLatSymMatElements}, and with $t$ as in \cref{fig:fig2}a2.

At $E = \mathcal{E}_1 = \sqrt{2}$, we have 
\begin{equation}
a(\mathcal{E}_1) = c(\mathcal{E}_1) = \mathcal{E}_1 \,.
\end{equation}
The non-linear eigenvalue problem \cref{eq:nlEVP} thus reads
\begin{equation}
\begin{pmatrix}
\mathcal{E}_1 & 1/\sqrt{2} + t e^{-ik} \\
1/\sqrt{2} + t e^{ik} & \mathcal{E}_1
\end{pmatrix} \Psi_S = \mathcal{E}_1 \Psi_S
\end{equation}
which is nothing but the eigenvalue equation for the edge states of the SSH model with intra-cell hopping $s:= 1/\sqrt{2}$ and with energy $E = \mathcal{E}_1$ instead of $E=0$.

The above leads to a general construction principle for a one-dimensional approximate topological insulator.
Given a Hamiltonian $H$ with a partial latent symmetry at $E=\mathcal{E}$ of two sites $S=\{u,v\}$, we build a one-dimensional lattice by taking $H$ as a unit cell, and with inter-cell hopping $t$ between sites $u,v$ of neighbouring unit cells.
In complete similarity to what we have done for \cref{fig:fig2}a2, we further add identical on-site potentials $\delta$ on all sites $u,v$, such that the diagonal elements of the ISR at energy $\mathcal{E}$ are identically given by $\mathcal{E}$.
For large enough $t$, the system then hosts two edge modes at energy $\mathcal{E}$.

\subsection*{Graph search}
For our statistics on the number of geometric and partial latent symmetries, we relied on the nauty (Not An AuTomorphism, Yes?) \cite{McKay2014JoSC6094PracticalGraphIsomorphismII} suite to generate all 11,989,733 simple connected graphs with up to $10$ vertices.
Each of these graphs corresponds to a symmetric adjacency matrix $A=A^T$, that can be identified with a tight-binding Hamiltonian with identical vanishing on-site potential (diagonal entries) and with hoppings of unity (two sites are connected) or zero (not connected).

We then checked each of these matrices individually for their symmetries.
A geometric symmetry corresponds to the existence of a permutation matrix $P$ commuting with $A$.
In graph-theory, such permutation symmetries are known as graph automorphisms and can be efficiently found using graph-theoretic algorithms.
In particular, the nauty suite is able to check if a given graph has such symmetries or not.
To verify whether the matrix $A$ has a partial latent symmetry is more difficult.
We diagonalized the matrix $A$ and subsequently computed the ISRs over all possible subsets of sites $S$, with the size of the subset ranging from $2$ to $N-1$, where $N$ is the dimension of the matrix $A$. 
Subsequently, we evaluated these ISRs at all eigenvalues individually and analyzed whether there is a permutation matrix $P$ that commutes with it.
\subsection*{(Nested) Wilson loops}
Both the Zak phase and the nested Wilson loop invariants require the calculation of Wilson loops. For a $D$-dimensional system, there will be $D$ momenta.
We consider here the case where the Wilson loop is taken along $k_1$. In this case, it is defined as
\begin{equation}
\mathcal{W}(k_2\dots k_D) = M^{(N)} M^{(N-1)}\cdots M^{(1)},
\end{equation}
where the matrices $M$ are given by 
\begin{equation}
M^{(i)}_{mn}= \braket{u_m(k_{1,i}\dots k_D)}{u_n(k_{1,i+1}\dots k_D)}\,.
\end{equation}
The indices $n$ range over the occupied subspace and the discretized momentum is given by $k_j = 2\pi j/N$. Since $\mathcal{W}$ is a unitary operator, we have
\begin{equation}
\mathcal{W}(k_2\dots k_D) \ket{\nu_j(k_2\dots k_D)} = e^{i \theta_j(k_2\dots k_D)} \ket{\nu_j(k_2\dots k_D)}.
\end{equation}
For one-dimensional systems, this yields
\begin{equation}
\mathcal{W} \ket{\nu_j} = e^{i\theta_j} \ket{\nu_j},
\end{equation}
from which the Zak phase, $\theta$, may be obtained:
\begin{equation}
\theta = \sum_j\theta_j.
\end{equation}
For two dimensions, one has
\begin{equation}
\mathcal{W}(k_2) \ket{\nu_j(k_2)} = e^{i\theta_j(k_2)} \ket{\nu_j(k_2)}.
\end{equation}
The nested Wilson loop is then obtained by taking the Wilson loop using the states $\ket{\nu_j}$.
\begin{equation}
\mathcal{W}^{\nu_{k_1}} = F_{\nu_{k_1}}^{(N)} F_{\nu_{k_1}}^{(N-1)} \cdots F_{\nu_{k_1}}^{(1)},
\end{equation}
where
\begin{equation}
[F_{\nu_{k_1}}^{(i)}]_{mn}= \braket{\nu_m(k_{2,i})}{\nu_n(k_{2,i+1})},
\end{equation}
Finally, the eigenvalues, $\theta_j^{\nu_{k_1}}$, of the nested Wilson loop
\begin{equation}
\mathcal{W}^{\nu_{k_1}} \ket{\nu_j^{\nu_{k_1}}} = e^{i \theta_j^{\nu_{k_1}}} \ket{\nu_j^{\nu_{k_1}}},
\end{equation}
are the topological invariants for the 2D systems \cite{benalcazarQuantizedElectricMultipole2017}.\\

\subsection*{Experimental Details}
The experimental realization of the Hermitian tight-binding Hamiltonian is achieved using a lumped-element topolectrical circuit network \cite{Lee2018CP11TopolectricalCircuits}. The circuit is constructed on a double layer printed circuit board (PCB) using FR4 substrate with a $1$~oz copper finish on both top and bottom layers.
To map the theoretical model onto the circuit system, each lattice site $n$ is represent by an electrical node, and the tight-binding hopping parameters are replaced by high-precision capacitors.
Specifically, the intra and inter-cell hoppings are implemented using capacitors $C_v=100$ pF (Murata GRM0335C1E101FA01), and $C_t=820$ pF (Murata GRM0335C1E821FA01), respectively.

To ensure that the circuit Laplacian corresponds to the desired Hamiltonian, each node $n$ is grounded via an inductor $L$ in series with a parasitic resistance $R_L$. The effective on-site admittance for each port is governed by the total nodal capacitance $C_{a,a}=2C_v+ 2C_t$, ensuring the diagonal terms of the circuit Laplacian offset the off-diagonal ones. The inductors are low-loss wired-wound components (CY105-100k) with a nominal inductance of $L=10 \mu H$ and intrinsic series resistance $R_L \approx 1 \Omega$. 

The electrical response of the circuit is governed by Kirchhoff's current law. In the frequency domain, the relationship between the external current input $\mathbf{I}(\omega)$ and the nodal voltages $\mathbf{V}(\omega)$ is defined by the circuit admittance matrix $\mathbf{J}(\omega)$ as $\mathbf{I}=\mathbf{J} \mathbf{V}$. For our Hermitian model, the admittance matrix is formulated as
\begin{equation}
    \mathbf{J}=i\omega C_{a,a} \mathbb{I}-i\omega C_{v,t} \mathbf{H}_p +\frac{1}{i\omega L+R_L} \mathbb{I},
    \label{AdmittanceMatrixFormula}
\end{equation}
where $\mathbb{I}$ is the identity matrix, $\mathbf{H}_p$ is the Bloch Hamiltonian of our model, and $C_{v,t}$ acts as the scaling capacitance for the hopping terms. In the ideal condition ($R_L \to 0$), the resonance condition of the circuit, where no external current flows ($\mathbf{I}=0$). By factoring out the common terms, the admittance matrix can be rearranged as
\begin{equation}
    \mathbf{J}= i\omega L \left( C_{a,a} \mathbb{I}-C_{v,t} \mathbf{H}_p \right)-\frac{1}{i\omega}.
\end{equation}
Setting $\mathbf{J} \mathbf{V}=0$ yields the eigenvalues problem for the resonance frequencies
\begin{equation}
    \left[ (C_{a,a} \mathbb{I} -C_{v,t} \mathbf{H}_p)L  \right] \mathbf{V}=\frac{1}{\omega^2} \mathbf{V}.
\end{equation}
This allows us to establish a direct mapping between the Hamiltonian eigenvalues $\lambda$ and the driving frequency $\omega$:
\begin{equation}
    \mathbf{H}_p \mathbf{V}=\left( \frac{\omega^2 L C_{a,a}-1}{\omega^2 LC_{v,t}} \right) \mathbf{V}=\lambda \mathbf{V}
\end{equation}
where $\lambda(\omega)=(\omega^2LC_{a,a}-1)/ \omega^2 LC_{v,t}$ serves as the frequency dependent spectral parameter.

To probe the topological characteristics, we measured the single port impedance $Z_{a,a}$ by using a Rohde \& Schwarz Vector Network Analyzer (ZNL20) connected via SMA ports. The impedance matrix $\mathbf{Z}$ is defined as the inverse of the admittance matrix $\mathbf{Z}=\mathbf{J}^{-1}$. Utilizing the spectral decomposition of $\mathbf{H}_p$ in the Hermitian system, we can express the full impedance matrix as
\begin{equation}
    \mathbf{Z}= \left( i\omega C_{a,a} \mathbb{I}-i\omega C_{v,t} \mathbf{H}_p+\frac{1}{i\omega L+R_L} \mathbb{I} \right)^{-1} \,.
\end{equation}
In the presence of a finite parasitic resistance $R_L$, the inductive term is modified as $(i \omega L+R_L)^{-1} \approx (i \omega L)^{-1}+ R_L/(\omega^2 L^2)$ for $\omega L \gg R_L$. By substituting this into the spectral expansion, the single port impedance measured at node $a$ is given by
\begin{equation}
    Z_{a,a}(\omega) =\frac{i}{\omega C_{v,t}} \sum_{n=1}^N \frac{|\psi_{n,a}|^2}{\lambda_n-\lambda(\omega)+i\gamma(\omega)} \,,
\end{equation}
where $\psi_{n,a}=\braket{a|v_n}$ is the amplitude of the $n$-th eigenmode at site $a$, and $\gamma(\omega)= R_L / (\omega^3 L^2 C_{v,t})$ is the effetive damping coefficient. The introduction of $\gamma(\omega)$ results in a Lorentzian broadening of the impedance peaks, preventing the divergence of $Z_{a,a}$ at the resonance condition $\lambda(\omega)=\lambda_n$\cite{Olekhno2020NC111436TopologicalEdgeStatesInteracting}.
The spatial profile of these peaks across different ports $a$ allows for the direct observation of the topological states's localization, corresponding to the local density of states $|\psi_{n,a}|^2$. Despite the peak broadening and finite height imposed by $R_L$, the relative magnitude of $Z_{a,a}$ remains a high fidelity probe for mapping the eigenmode wavefunctions.

\vspace{1cm}
\section{Author contributions}
The work was conceived jointly by L.E., W.G., and M.R.
X.C. planned and conducted the experiments.
L.E. performed the theoretical analysis of the presented models in terms of their topological properties.
M.R. conducted the graph-theoretical analysis.
W.G., M.R., and C.M.S. supervised the work.
All the authors discussed the results and contributed to this work.
The manuscript was mainly written by L.E. and M.R., with additional input from all the other authors.

\end{document}


\title{Supplemental Material for: Ubiquitous topological materials protected by hidden partial symmetries}

\author{Lumen Eek\,\orcidlink{0009-0009-1233-4378}}
\affiliation{Institute of Theoretical Physics, Utrecht University, Utrecht, 3584 CC, Netherlands}

\author{Xuelong Chen\,\orcidlink{0009-0002-3885-5402}}
\affiliation{Eastern Institute for Advanced Study, Eastern Institute of Technology, Ningbo, China}
\affiliation{Department of Electrical and Electronic Engineering, The Hong Kong Polytechnic University, Hong Kong, China}

\author{Malte Röntgen\,\orcidlink{0000-0001-7784-8104}}
\email{mroentgen@eitech.edu.cn}
\affiliation{Eastern Institute for Advanced Study, Eastern Institute of Technology, Ningbo, China}

\author{Wenlong Gao\,\orcidlink{0000-0002-3446-7157}}
\email{wgao@eitech.edu.cn}
\affiliation{Eastern Institute for Advanced Study, Eastern Institute of Technology, Ningbo, China}

\author{Cristiane Morais Smith\,\orcidlink{0000-0002-4190-3893}}
\affiliation{Institute of Theoretical Physics, Utrecht University, Utrecht, 3584 CC, Netherlands}

\maketitle

\tableofcontents

\newpage
\section{Additional details on the different types of SSH model}
\label{sec:latent}
\begin{figure}
    \centering
    \includegraphics[width=\linewidth]{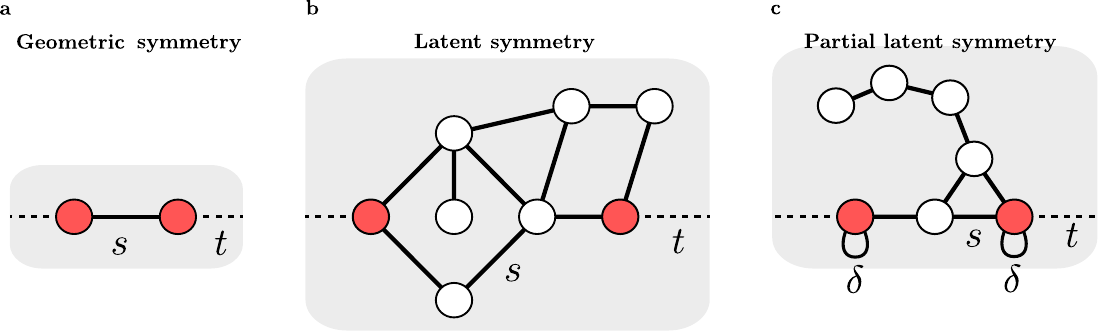}
    \caption{\textbf{SSH-type models.} \textbf{a}, SSH model with intra- and intercell hopping $s$ and $t$ respectively. \textbf{b}, Latent SSH model; all intracell hoppings are equal to $s$ while the intercell hopping is equal to $t$. \textbf{c}, Partial Latent SSH model; intracell hoppings are equal to $s$ while the intercell hopping is equal to $t$. Additionally the red sites have an onsite potential $\delta$.}
    \label{fig:models}
\end{figure}

In this section, we provide some more information on the different type of symmetry-protected topological models treated in the main text, additionally we have repeated them in Fig.~\ref{fig:models}.

\subsection{Classic SSH model}
\begin{figure}[H]
    \centering
    \includegraphics[width=0.8\linewidth]{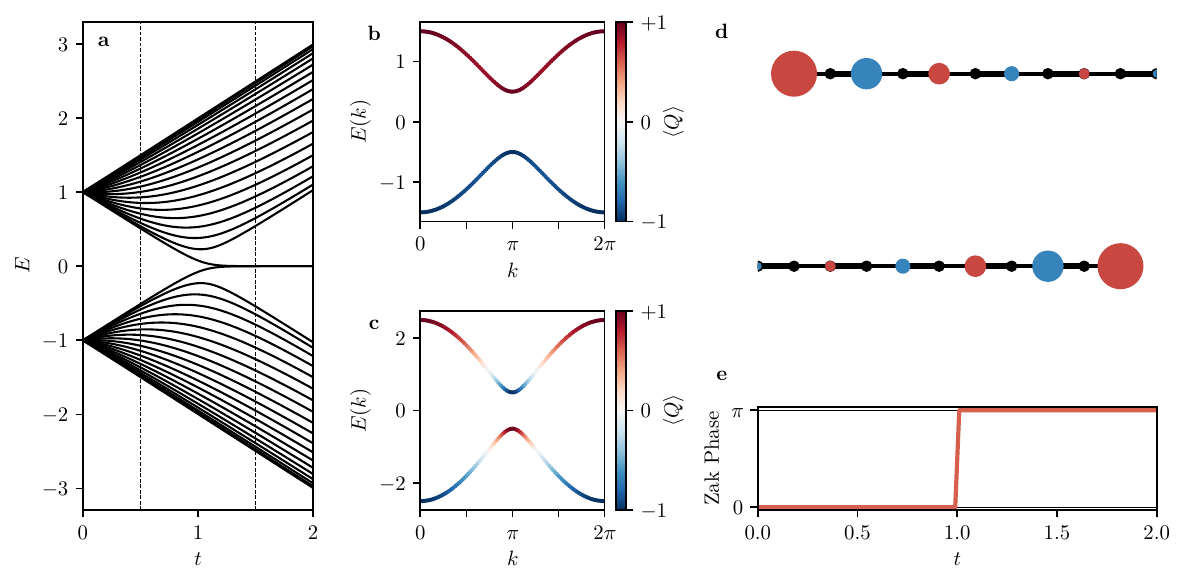}
    \caption{
    \textbf{Topological properties of the SSH model. }
    \textbf{a}, Energy spectrum under open boundary conditions as a function of the hopping ratio $t$, showing a gap closing at $|t|=1$ and the emergence of in-gap edge states for $|t|>1$. 
    \textbf{b,c}, Bulk band structure under periodic boundary conditions for $t=1/2$ (\textbf{b}) and $t=3/2$ (\textbf{c}), with colours indicating the inversion expectation value $\langle Q \rangle$. A band inversion occurs in the topological phase. 
    \textbf{d}, Spatial profile of the edge states for $t=3/2$, demonstrating localization at opposite ends of the chain. Size of the circles indicate wavefunction amplitude, while color represents opposite sign.
    \textbf{e}, Zak phase as a function of $t$, showing quantization to $0$ (trivial) and $\pi$ (topological). All hoppings and energies in this figure are in units of the intracell hopping, which is set to $1$.
}\label{fig:SSH}
\end{figure}
To build intuition for our results, we first revisit the paradigmatic Su–Schrieffer–Heeger (SSH) model. Its Bloch Hamiltonian is given by
\begin{equation} \label{eq:OriginalSSH}
    H_{\mathrm{SSH}}(s, k) = \begin{pmatrix}
        0 & s + t e^{i k} \\
        s + t e^{-i k} & 0
    \end{pmatrix}\,,
\end{equation}
aditionally, the unit cell is presented in Fig.~\ref{fig:models}a. In the following, we set the intracell hopping amplitude $s$ to unity.
The system undergoes a gap closing at $|t|=1$, signalling a topological phase transition, as reflected in the open-boundary (OBC) spectrum shown in Fig.~\ref{fig:SSH}a. For $|t|>1$, the system realizes a topological phase characterized by a pair of in-gap edge states localized at opposite ends of the chain. Their spatial profiles for $t=3/2$ are displayed in Fig.~\ref{fig:SSH}d. Note that the color indicates the sign of the wavefunction and not the sublattice. 
The corresponding periodic-boundary (PBC) spectra for $t=1/2$ and $t=3/2$ are shown in Fig.~\ref{fig:SSH}b,c, respectively (see vertical dashed lines in Fig.~\ref{fig:SSH}a). The bands are coloured according to their inversion expectation value $\langle Q \rangle = \langle u_{n k} | \sigma_x | u_{n k} \rangle$, where $\sigma_x$ acts as the inversion operator. In the topological phase, a clear band inversion is observed (Fig.~\ref{fig:SSH}c).
Finally, the Zak phase, shown in Fig.~\ref{fig:SSH}e, exhibits the expected quantization to $0$ or $\pi$, distinguishing the trivial and topological phases.

\subsection{Latent SSH model}

Now, we consider an example of a latent SSH model, following Ref.~\cite{Rontgen2024PRB110035106TopologicalStatesProtectedHidden}.
Its Bloch Hamiltonian is given by
\begin{equation}
    H_{LSSH}(k) = 
\begin{pmatrix}
0 & te^{ik} & 1 & 0 & 1 & 0 & 0 & 0 \\
te^{-ik} & 0 & 0 & 0 & 0 & 1 & 0 & 1 \\
1 & 0 & 0 & 1 & 0 & 1 & 1 & 0 \\
0 & 0 & 1 & 0 & 0 & 0 & 0 & 0 \\
1 & 0 & 0 & 0 & 0 & 1 & 0 & 0 \\
0 & 1 & 1 & 0 & 1 & 0 & 1 & 0 \\
0 & 0 & 1 & 0 & 0 & 1 & 0 & 1 \\
0 & 1 & 0 & 0 & 0 & 0 & 1 & 0
\end{pmatrix}\,,
\end{equation}

aditionally, the unit cell is presented in Fig.~\ref{fig:models}b. Performing the ISR on sites $1$ and $2$ yields an effective, energy-dependent Bloch Hamiltonian
\begin{equation} \label{eq:latentSSH}
    \mathcal{R}_{S=\{1,2\}}[H_{\mathrm{LSSH}}(k),E] =
    \begin{pmatrix}
        a(E) & b(E) + t e^{i k} \\
        b(E) + t e^{-i k} & a(E)
    \end{pmatrix},
\end{equation}
with
\begin{align}
    a(E) &= \frac{2 E^3 - 2 E^2 - 2 E}{(E^2 - 2 E - 1)(E^2 + E - 1)},\\
    b(E) &= \frac{2 E^2}{E^4 - E^3 - 4 E^2 + E + 1}.
\end{align}
Comparing \cref{eq:latentSSH} with \cref{eq:OriginalSSH}, we see that 
\begin{equation} \label{eq:sshMapping}
     \mathcal{R}_{S=\{1,2\}}[H_{\mathrm{LSSH}}(k),E] = H_{\mathrm{SSH}}(b(E), k) + a(E) I\,.
\end{equation}
Generally speaking, a latent SSH model is one whose ISR fulfils \cref{eq:sshMapping}; that is, the ISR maps the model onto an energy-dependent SSH model (with additional on-site potential $a(E)$), for $\textbf{all}$ energies. 
\begin{figure}
\centering
    \includegraphics[width=0.8\linewidth]{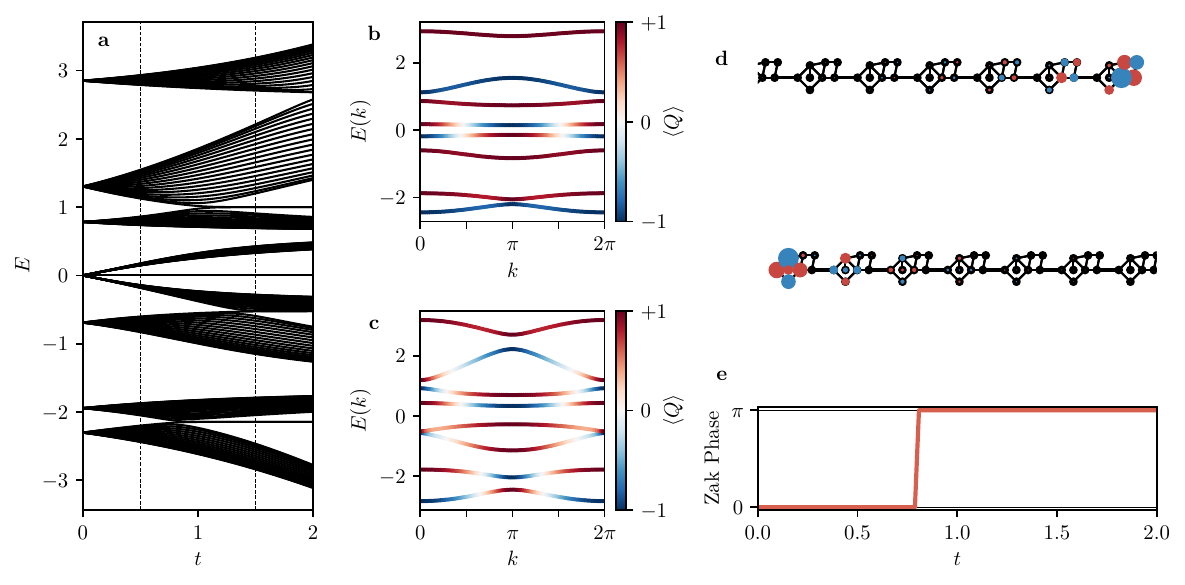}
    \caption{\textbf{Topological properties of the Latent SSH model. } \textbf{a},  Open-boundary spectrum showing in-gap edge states at the transition energies $E_i$.  \textbf{b,c} Bulk band structure under periodic boundary conditions for $t=1/2$ (\textbf{b}) and $t=3/2$ (\textbf{c}), with colours indicating the inversion expectation value $\langle Q \rangle$. A band inversion indicates a topological gap. \textbf{d}, Spatial profile of an edge state, illustrating localization at the (right and left) chain ends. Size of the circles indicate wavefunction amplitude, while color represents the sign of the real-valued wavefunction. \textbf{e}, Zak phase of the lowest gap, computed from the effective two-band ISR Hamiltonian, quantized to $0$ or $\pi$, distinguishing trivial and topological phases. All hoppings and energies in this figure are in units of the intracell hopping, which is set to $1$.
}\label{fig:LSSH}
\end{figure}
The algebraic mapping \cref{eq:sshMapping} is quite convenient.
In particular, we can use it to write the eigenvalue equation $\mathcal{R}_{S=\{1,2\}}[H_{\mathrm{LSSH}}(k),E] \Psi(k) = E \Psi(k)$ for the latent SSH model as
\begin{equation}
H_{\mathrm{SSH}}\left(b(E), k\right) \Psi(k) = (E-a(E)) \Psi(k)\,.
\end{equation}
The topological transition of the original SSH model occurs at eigenvalue zero, and for $|s| = |t|$.
Consequently, our latent SSH model has multiple topological transitions, each occurring at an energy $E_i$ satisfying $E_i - a(E_i)= 0$. The corresponding intra-cell coupling $b(E_i)$ defines the critical hopping strength: the $i$-th transition takes place when $|b(E_i)| = |t|$.

The spectrum under open boundary conditions (OBC) of the full Hamiltonian, $H_{LSSH}(k)$, shows in-gap edge states appearing at the energies $E_i$, localized at opposite ends of the chain [Fig.~\ref{fig:LSSH}a,\ref{fig:LSSH}d]. The periodic-boundary (PBC) bulk bands [Figs.~\ref{fig:LSSH}b,\ref{fig:LSSH}c] are colored according to the inversion expectation value
\begin{equation}
\langle Q \rangle = \langle u_{nk} | Q | u_{nk} \rangle,
\end{equation}
where the inversion operator $Q$ is given by
\begin{equation}
Q =
\begin{pmatrix}
0 & 1 & 0 & 0 & 0 & 0 & 0 & 0 \\
1 & 0 & 0 & 0 & 0 & 0 & 0 & 0 \\
0 & 0 & \frac{1}{2} & 0 & -\frac{1}{2} & \frac{1}{2} & 0 & \frac{1}{2} \\
0 & 0 & 0 & \frac{1}{2} & 0 & -\frac{1}{2} & \frac{1}{2} & \frac{1}{2} \\
0 & 0 & -\frac{1}{2} & 0 & \frac{1}{2} & \frac{1}{2} & 0 & \frac{1}{2} \\
0 & 0 & \frac{1}{2} & -\frac{1}{2} & \frac{1}{2} & 0 & \frac{1}{2} & 0 \\
0 & 0 & 0 & \frac{1}{2} & 0 & \frac{1}{2} & \frac{1}{2} & -\frac{1}{2} \\
0 & 0 & \frac{1}{2} & \frac{1}{2} & \frac{1}{2} & 0 & -\frac{1}{2} & 0
\end{pmatrix}.
\end{equation}
We note that $Q$ fulfils $Q^2 = I$, $Q^T = Q^{-1}$, as well as $Q H_{LSSH}(k) Q = H_{LSSH}(-k)$.
Such an operator $Q$ always exists for a latent SSH system and can be derived using projection operators, as is explained in more detail in Appendix D of Ref.~\cite{Rontgen2024PRB110035106TopologicalStatesProtectedHidden}, as well as Sec. IV of the Supplemental Material of Ref.~\cite{Rontgen2021PRL126180601LatentSymmetryInducedDegeneracies}.

In the topological phase, multiple band inversions can be observed in Fig.~\ref{fig:LSSH}c. In particular for the lowest gap (around $E=-2$) an inversion can be seen. The Zak phase [Fig.~\ref{fig:LSSH}e] corresponding to this gap, is quantized to $0$ or $\pi$, corresponding to trivial and topological phases. Additionally, other band inversions can be observed, for example around $E=0$ and $E=1$. The bands around $E=0$ are inverted for any choice of $t$, as shown in Fig.~\ref{fig:LSSH}c. 

This demonstrates that the latent SSH model can host multiple independent topological transitions, all explained within the effective two-band ISR description.

\subsection{Partial latent SSH model}

\begin{figure}
    \centering
    \includegraphics[width=0.8\linewidth]{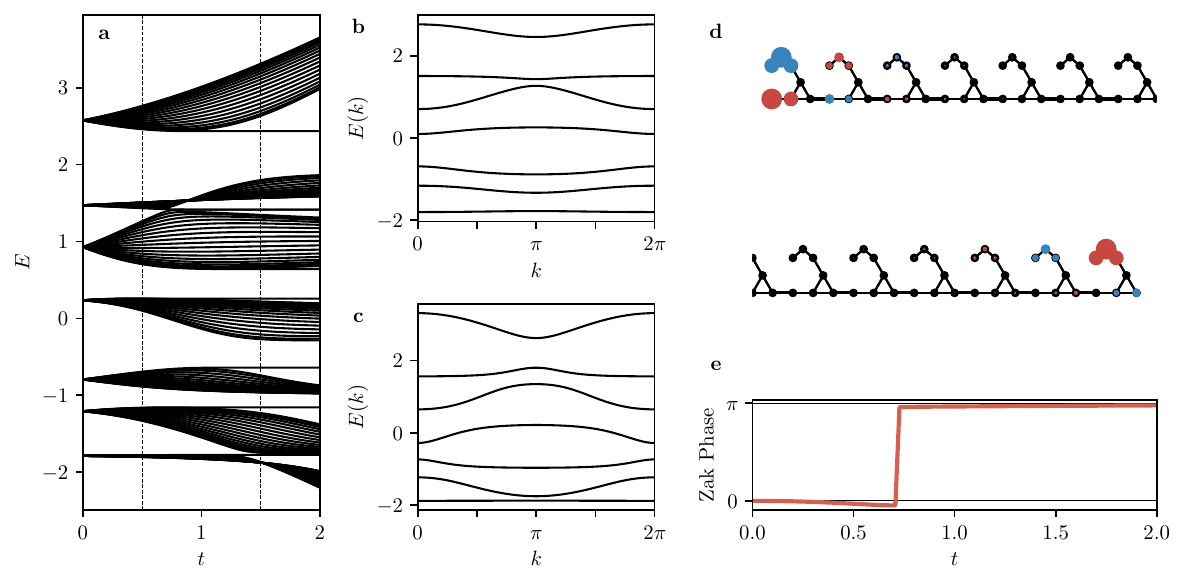}
    \caption{\textbf{Topological properties of the Partial Latent SSH model. } 
    \textbf{a}, Open-boundary spectrum showing in-gap edge states at energy $\varepsilon = \sqrt{2}$.
    \textbf{b},\textbf{c} Bulk band structures for inter-cell hopping $t=1/2$ and $t=3/2$, respectively. 
    \textbf{d}, Topological in-gap end states at $E=\sqrt{2}$ for $t=3/2$, shown at both ends of the chain.
    \textbf{e}, Zak phase of the gap around $E=\varepsilon$, which is approximately quantized and jumps when $|t|=|b(\varepsilon)|$. All hoppings and energies in this figure are in units of the intracell hopping, which is set to $1$.
    }
    \label{fig:PLSSH}
\end{figure}

Finally, we turn our attention to the partial latent SSH model, treated in the main text, the unit cell is repeated in Fig.~\ref{fig:models}c.
A finite size realization consisting of four unit cells is depicted in \cref{fig:partialSSHBoundaries}~a.
Performing a Fourier transformation (of an infinitely extenddd lattice), we obtain the Bloch Hamiltonian of the full system as
\begin{equation}\label{eq:PLSSH}
H_{PLSSH}(k) =
\begin{pmatrix}
\delta & t e^{ik} & 1 & 0 & 0 & 0 & 0 \\
te^{-ik} & \delta & 1 & 1 & 0 & 0 & 0 \\
1 & 1 & 0 & 1 & 0 & 0 & 0 \\
0 & 1 & 1 & 0 & 1 & 0 & 0 \\
0 & 0 & 0 & 1 & 0 & 1 & 0 \\
0 & 0 & 0 & 0 & 1 & 0 & 1 \\
0 & 0 & 0 & 0 & 0 & 1 & 0
\end{pmatrix}
\end{equation}

Upon performing the ISR on sites $1$ and $2$, we obtain the effective (energy dependent) Bloch Hamiltonian
\begin{equation}\label{eq:reducedPLSSH}
    \mathcal{R}_{S=\{1,2\}}[H_{PLSSH}(k),E] =
    \begin{pmatrix}
        \tilde{a}(E) & b(E) + t e^{i k} \\
        b(E) + t e^{-i k} & \tilde{c}(E)
    \end{pmatrix}
    \,.
\end{equation}
The expression for $\tilde{a}(E)$, $b(E)$, and $\tilde{c}(E)$ are given in the Methods section of the main text. For completeness we repeat them here;
\begin{align} \label{eq:aExp}
    \tilde{a}(E) &= a(E) + \delta = \frac{E ^4-3 E ^2+1}{E ^5-4 E ^3+3 E } + \delta \,,\\
    b(E) &= \frac{E ^3-3 E +1}{(E -1) E  \left(E ^2-3\right)} \,,\\
    \tilde{c}(E) &= c(E) + \delta = \frac{2 E ^3-5 E +1}{(E -1) E  \left(E ^2-3\right)}+\delta \,.
\end{align}

The partial latent SSH model generally has only one topological transition.
To see this, we need to realize that a necessary condition is that there is an energy $\varepsilon$ for which $a(\varepsilon) = c(\varepsilon)$.
At this energy, the eigenvalue equation becomes
\begin{equation} \label{eq:partialSshMapping}
H_{\mathrm{SSH}}\left(b(\varepsilon), k\right) \Psi(k) = (\varepsilon-a(\varepsilon) - \delta) \Psi(k)\,.
\end{equation}
In order to have a topological transition---which, we remind the reader, occurs in the classic SSH model at energy $E = 0$---, we thus additionally need $a(\varepsilon) + \delta = \varepsilon$.
Since $\epsilon$ can only take a usually small number of finite values, $\delta$ usually---though not always, as we shall see later when discussing the partial latent BBH model---needs to fulfil this.
For the partial latent SSH model discussed here, we have $\delta = 1/\sqrt{2}$.
Topological edge states at energy $E = \varepsilon$ then emerge for $|t| > |b(\varepsilon)|$, i.e., when the inter-cell coupling is stronger than the intra-cell coupling.  

A comparison with the latent SSH model reveals that while the latent SSH model reduces to an effective SSH model at all energies, a partial latent SSH model only reduces to an SSH model at a specific energy $\varepsilon$.

Figure~\ref{fig:PLSSH}a shows the open-boundary spectrum of Eq.~\eqref{eq:PLSSH} as a function of inter-cell hopping. Multiple edge states are visible, but only those at $E = \varepsilon = \sqrt{2}$ are pinned, consistent with the partial latent symmetry. Bulk bands for $t=1/2$ and $t=3/2$ are depicted in Fig.~\ref{fig:PLSSH}b,c, and the topological in-gap end states at $E = \sqrt{2}$ for $t = 3/2$ are shown in Fig.~\ref{fig:PLSSH}d. Due to the absence of a symmetry at arbitrary energies, no symmetry matrix $Q$ exists to extract inversion eigenvalues. Furthermore, by lack of an exact protecting symmetry, the Zak phase of the topological gap is not quantized, but instead approximately quantized, as shown in Fig.~\ref{fig:PLSSH}e.

\subsubsection{Zak phase for $\mathcal{R}_{S=\{1,2\}}[H_{PLSSH}(k),\varepsilon]$}

A quantized Zak phase may be obtained by treating $\mathcal{R}_{S=\{1,2\}}[H_{PLSSH}(k),\varepsilon]$ as an effective Hamiltonian only for the end states at $E = \varepsilon$. Substituting $E = \varepsilon$ into Eq.~\eqref{eq:reducedPLSSH} yields
\begin{equation}
    \mathcal{R}_{S=\{1,2\}}[H_{PLSSH}(k),\varepsilon] = 
    \begin{pmatrix}
        \sqrt{2} & \frac{1}{\sqrt{2}} + t\,e^{i k} \\
        \frac{1}{\sqrt{2}} + t\,e^{-i k} & \sqrt{2}
    \end{pmatrix}.
\end{equation}

This is exactly the SSH model, with a quantized Zak phase and a topological phase transition occurring at $|t| = 1/\sqrt{2}$. This value matches the discontinuity of the Zak phase, as observed in Fig.~\ref{fig:PLSSH}e.

\subsubsection{Boundary effects}
\label{sec:edge}
\begin{figure}
	\centering
	\includegraphics[width=0.65\linewidth]{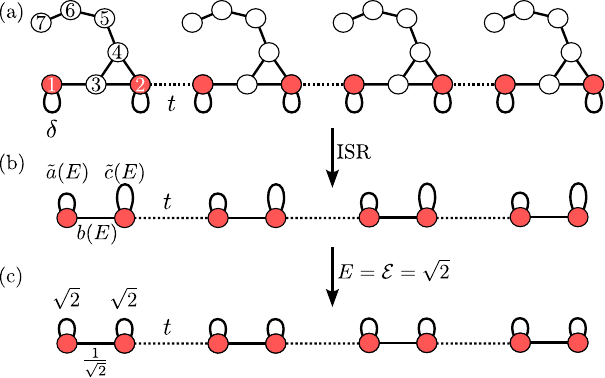}
	\caption{\textbf{A finite-size partial latent SSH model}.
    \textbf{a}, Small example setup with four unit cells, with sites in the first unit cell numbered according to \cref{eq:PLSSH}. 
    Intra-cell and inter-cell couplings are given by unity and $t$, respectively.
    \textbf{b}, ISR of the setup over the red sites, that is, sites $1$ and $2$ in each unit cell, yields a one-dimensional chain. Expressions for $\tilde{a}(E)$, $\tilde{c}(E)$, and $b(E)$ are given in \cref{eq:aExp}.
    \textbf{c}, At $E = \mathcal{E} = \sqrt{2}$, the chain becomes a SSH chain with identical on-site potential $\sqrt{2}$, intra-cell coupling $1/\sqrt{2}$, and inter-cell coupling $t$.
	}
	\label{fig:partialSSHBoundaries}
\end{figure}

To complement our discussion above, let us now investigate a finite partial latent SSH model and the consequences of its finite size in more detail.
In \cref{fig:partialSSHBoundaries}a, we show such a model, consisting of four unit cells.

There are two finite-size effects to take care of.
In principle---we shall see this phenomenon later---, for a finite system, the effective on-site potentials at the boundary may differ from those in the bulk.
For our partial latent SSH model, however, this is not the case, as can be seen in Figs.~\ref{fig:partialSSHBoundaries}b, \ref{fig:partialSSHBoundaries}c.

\begin{figure}
    \centering
    \includegraphics[width=0.5\linewidth]{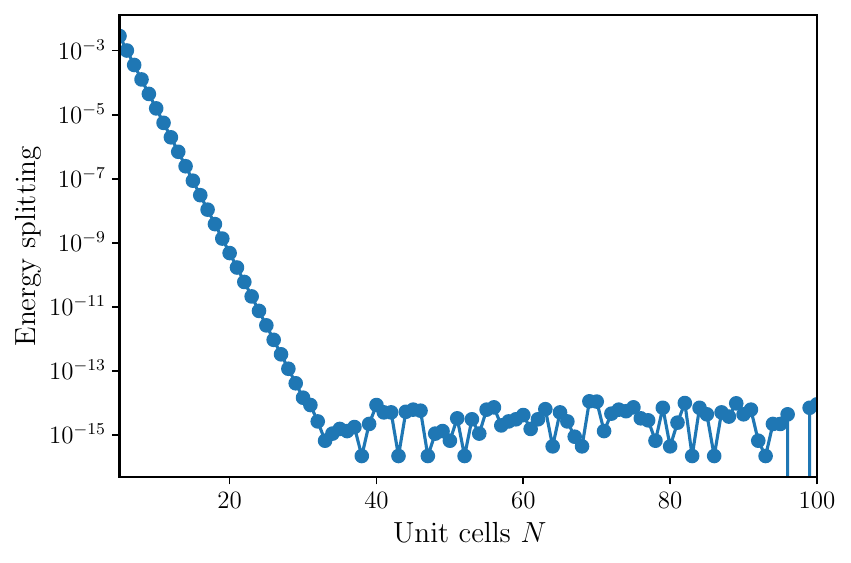}
    \caption{\textbf{Edge-state energy splitting for the partial Latent SSH model.} Splitting of the edge state energies for $t=2$, as a function of the chain length in number of unit cells $N$.}
    \label{fig:split}
\end{figure}

The second finite-size effect is energy-related.
To this end, let us explicitly write down the Hamiltonian of the finite-size chain after the ISR:
\begin{equation}
	\mathcal{R}_S(H_{PLSSH},E) = 
	\left(
	\begin{array}{cccccccc}
		\tilde{a}(E) & b(E) & 0 & 0 & 0 & 0 & 0 & 0 \\
		b(E) & \tilde{c}(E)& t & 0 & 0 & 0 & 0 & 0 \\
		0 & t & \tilde{a}(E) & b(E) & 0 & 0 & 0 & 0 \\
		0 & 0 & b(E) & \tilde{c}(E)& t & 0 & 0 & 0 \\
		0 & 0 & 0 & t & \tilde{a}(E) & b(E) & 0 & 0 \\
		0 & 0 & 0 & 0 & b(E) & \tilde{c}(E)& t & 0 \\
		0 & 0 & 0 & 0 & 0 & t & \ddots & \ddots \\
		0 & 0 & 0 & 0 & 0 & 0 & \ddots &  \ddots\\
	\end{array}
	\right) \,.
\end{equation}
We remind the reader that the eigenvalue equation for the ISR reads
\begin{equation}
	\mathcal{R}_S(H_{PLSSH},E) \psi = E \psi \,.
\end{equation}
Let us now assume that $E = \mathcal{E} = \sqrt{2}$ is an eigenvalue of the ISR (and thus also of $H_{PLSSH}$ \cite{Bunimovich2014IsospectralTransformationsNewApproach}).
Since $\tilde{a}(E) = \tilde{c}(E) = \mathcal{E}$, the corresponding eigenvector $\psi$ is \emph{identical} to the eigenvector of a classic SSH model\footnote{with inter-cell coupling $t$ and intra-cell coupling $s = b(\mathcal{E})$.} with energy $0$; that is, an exponentially localized edge state.
Since $\psi$ is an eigenvector of the ISR, there exists an eigenvector $\Psi$ of the full Hamiltonian $H_{PLSSH}$ at the same energy $E = \mathcal{E}$; moreover, $\psi = \Psi_S$ is nothing but the $S$ components of $\Psi$.
From the derivation of the ISR (see main text), it follows that the components of this eigenvector on the remaining sites $\sbar$ can be obtained through
\begin{equation}
	\Psi_{\sbar} = \left( H_{\overline{SS}} - E I \right)^{-1} H_{\sbar S} \Psi_S \,.
\end{equation}
Using the structure of our partial latent SSH model---in particular, the fact that $H_{\overline{SS}}$ is block-diagonal, and that a given site $s \in S$ only couples to one of these blocks---,it can be easily shown that, since $\Psi_S$ is exponentially decaying, so is $\Psi_{\sbar}$.
In reality, though, a finite-size chain will not have an eigenvalue exactly equal to $\mathcal{E}$; this is analogous to a finite SSH chain, which will not have an eigenvalue that is exactly $0$.
For both chains, there exist eigenvalues that are approximately equal to these values ($\mathcal{E}$ for the partial latent SSH model, and $0$ for the classic SSH model); the longer the chain, the smaller the deviation.
For our partial latent SSH model, this deviation in energy needs special attention; since $E \approx \mathcal{E}$, we only have $\tilde{a}(E) \approx \tilde{c}(E)$; thus, a finite-size partial latent SSH chain is similar to a classic SSH model with an on-site potential imbalance $\Delta = \tilde{a}(E) -  \tilde{c}(E)$.
In our system, however, this imbalance is not strong and can be neglected for sufficiently long chains, as can be seen in Fig.~\ref{fig:split}. Already at a chain length of $N=20$ unit cells the edge states are pinned together to $E=\mathcal{E}$ with a deviation of less than $10^{-9}$.

\section{Additional details on the Hermitian latent BBH model}

\begin{figure}
    \centering
    \includegraphics[width=0.6\linewidth]{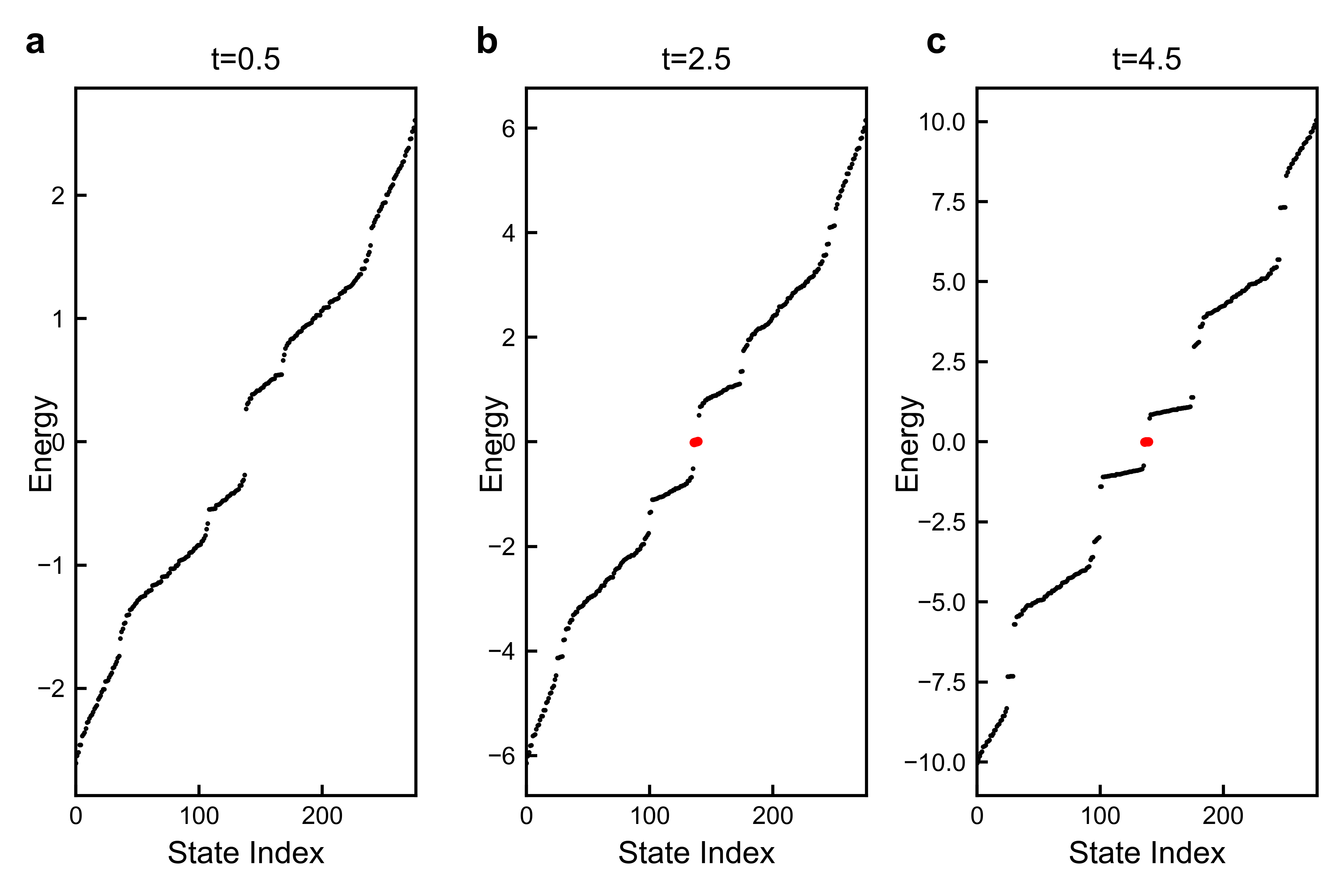}
    \caption{\textbf{Additional spectra of the partial latent BBH model.} OBC spectrum of the partial latent symmetric BBH model for \textbf{a}: $t=0.5$, \textbf{b}: $t=2.5$, \textbf{c}: $t=4.5$. In all systems we set $v=1$.
    }
    \label{fig:additionalBBH}
\end{figure}

The full Bloch-Hamiltonian of the BBH-type model in the main text is given by
\begin{equation}
H(k) = 
\begin{pmatrix}
    0 & v + t e^{-i k_y} & 0 & 0 & v & 0 & t & 0 \\
    v + t e^{i k_y} & 0 & v + t e^{-i k_x} & 0 & 0 & 0 & 0 & 0 \\
    0 & v + t e^{i k_x} & 0 & v + t e^{i k_y} & 0 & 0 & 0 & 0 \\
    0 & 0 & v + t e^{-i k_y} & 0 & 0 & v & 0 & t \\
    v & 0 & 0 & 0 & 0 & v & 0 & 0 \\
    0 & 0 & 0 & v & v & 0 & 0 & 0 \\
    t & 0 & 0 & 0 & 0 & 0 & 0 & t e^{-i k_x} \\
    0 & 0 & 0 & t & 0 & 0 & t e^{i k_x} & 0
\end{pmatrix}
\end{equation}
Performing the ISR, we reduce the original $8\times8$ Bloch Hamiltonian $H$ into a $4 \times 4$ effective Hamiltonian $H_{\mathrm{eff}}(E) = H_{SS} + H_{S\bar{S}}(EI - H_{\bar{S}\bar{S}})^{-1}H_{\bar{S} S}$, where $S$ denotes the retained lattice sites,
\begin{equation} \label{eq:BBHBlockEffective}
H_{\mathrm{eff}}(E,k) = \begin{pmatrix}
\varepsilon(E) & v+te^{-ik_{y}} & 0 & \kappa(E) \\
v+te^{ik_{y}} & 0 & v+te^{-ik_{x}} & 0 \\
0 & v+te^{ik_{x}} & 0 & v+te^{ik_{y}} \\
\kappa(E) & 0 & v+te^{-ik_{y}} & \varepsilon(E)
\end{pmatrix} \,.
\end{equation}
Here, the induced on-site energy $\varepsilon(E) = Ev^2/(E^2-v^2) + Et^2/(E^2-t^2)$ and the effective inter-cell coupling $\kappa(E)=v^3/(E^2-v^2) + t^3/(E^2-t^2)e^{\pm ik_{x}}$ represents the interaction with the removed auxiliary sites $\sbar$.
For $E=0$, $\varepsilon(0)=0$ and $\kappa(0)=-v -te^{\pm ik_{x}}$; this is the classical BBH model Hamiltonian. In Fig.~\ref{fig:additionalBBH} we depict the OBC spectrum of this model for parameters different than in the main text, for both the topological ($|t|>1)$ and trivial phase ($|t|<1$).

\subsection{Finite-size effects} \label{sec:BBHFiniteSize}
\begin{figure}
	\centering
	\includegraphics[width=0.7\linewidth]{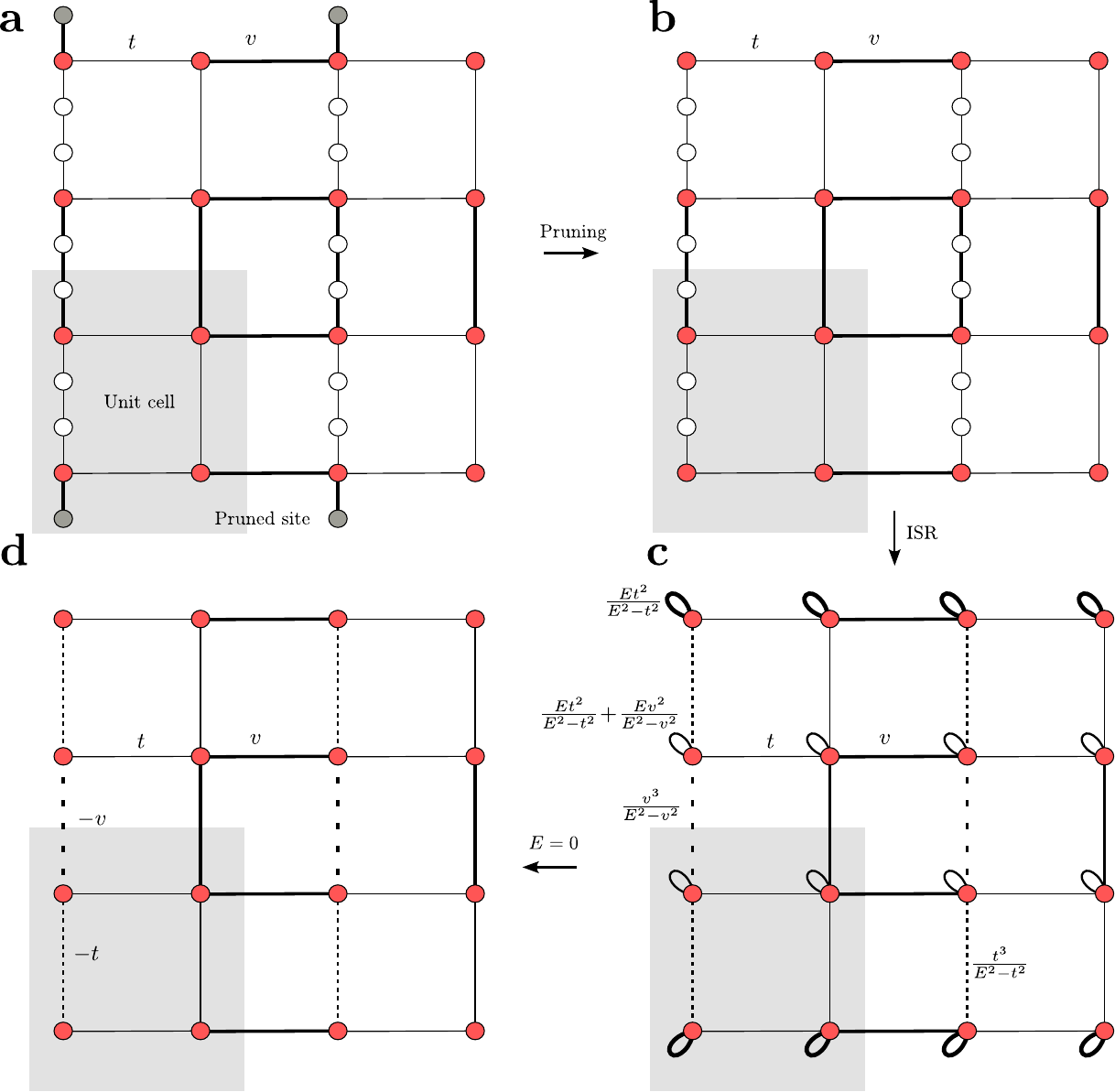}
	\caption{\textbf{A finite-size partial latent BBH model with $2\times{}2$ unit cells.}
    \textbf{a},
    Before pruning.
    \textbf{b},
    Pruning removes the upper and lower-most stubs.
    \textbf{c},
    The ISR performed upon all red sites.
	Without pruning, the on-site potential of the uppermost and lowermost sites would not be equal to $E  t^2/(E ^2-t^2)$, but to $E  t^2/(E ^2-t^2)+v^2/E$.
    \textbf{d},
    At $E=0$, the setup is equal to the classic BBH model.
	}
	\label{fig:BBHBoundaries}
\end{figure}

After the above analysis of the bulk Hamiltonian and its ISR, let us now consider a finite-size system, such as the one depicted in \cref{fig:BBHBoundaries}.
For each unit cell in the upper or lower end of the system, there is one stub site (marked in grey).
These stubs are problematic, as they lead to an ISR that diverges at $E=0$, that is, at precisely the energy we are interested in.
As a remedy, in everything that follows, we prune these stub sites (cf. \cref{fig:BBHBoundaries}~a,b).
The ISR of the pruned system is depicted in \cref{fig:BBHBoundaries}c.
It is well-defined at $E=0$, where all on-site (diagonal) terms vanish and where the system becomes the classic BBH model.

In complete analogy to the case of the partial latent SSH model, our finite-size system then behaves as the classic BBH model at $E=0$.
In particular, for large enough coupling $t$, corner states appear.
Similar to the case of the partial latent SSH model, the eigenvalue $E$ of these corner states will not be exactly equal to $0$, but slightly deviate from it.
As a consequence, these states do not ``see'' the BBH model, but one with slightly deviating couplings and on-site potentials.
For very large setups, these effects become negligible, similar to the 1D case.

\newpage
\subsection{Details on the experimental realization}

We realized a system containing $6\times 6 = 36$ unit cells, each containing $8$ sites.
After pruning (cf. \cref{sec:BBHFiniteSize}), our system thus consists of $36 \times 8 - 2\times 6 = 276$ sites.

In our experiment, the finite-size effects discussed in \cref{sec:BBHFiniteSize} are combined with inherent component disorder as well as parasitic losses.
Together, these lead to discrepancies among the measured impedance (representing density of states) at different corners.

\begin{figure}
    \centering
    \includegraphics[width=0.3\linewidth]{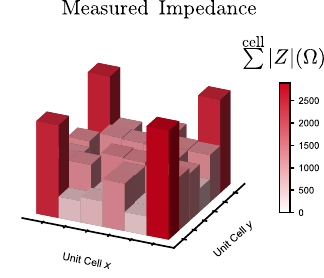}
    \caption{\textbf{Measured spatial distribution of the partial latent BBH circuit impedance at $f=1.248$~MHz.} The plot illustrates the localization of corner states and the observed amplitude asymmetry from experimental error.
    }
    \label{1.248MHz_Z}
\end{figure}

In Fig.~3 of the main text, we have shown the spatial distribution of the measured impedance at frequency ($1.259$~MHz).
A comparision of the measured impedance at the four corners against the measuring frequency ($1.248$~MHz) is shown in \cref{1.248MHz_Z}.

\newpage
\section{The non-Hermitian latent BBH model}

\subsection{Model}
The working principle behind partial latent symmetry topological insulators is not limited to Hermitian models. In the following section, we give an example of a non-Hermitian partial latent symmetry model.  
\begin{figure}[b]
    \centering
    \includegraphics[width=\linewidth]{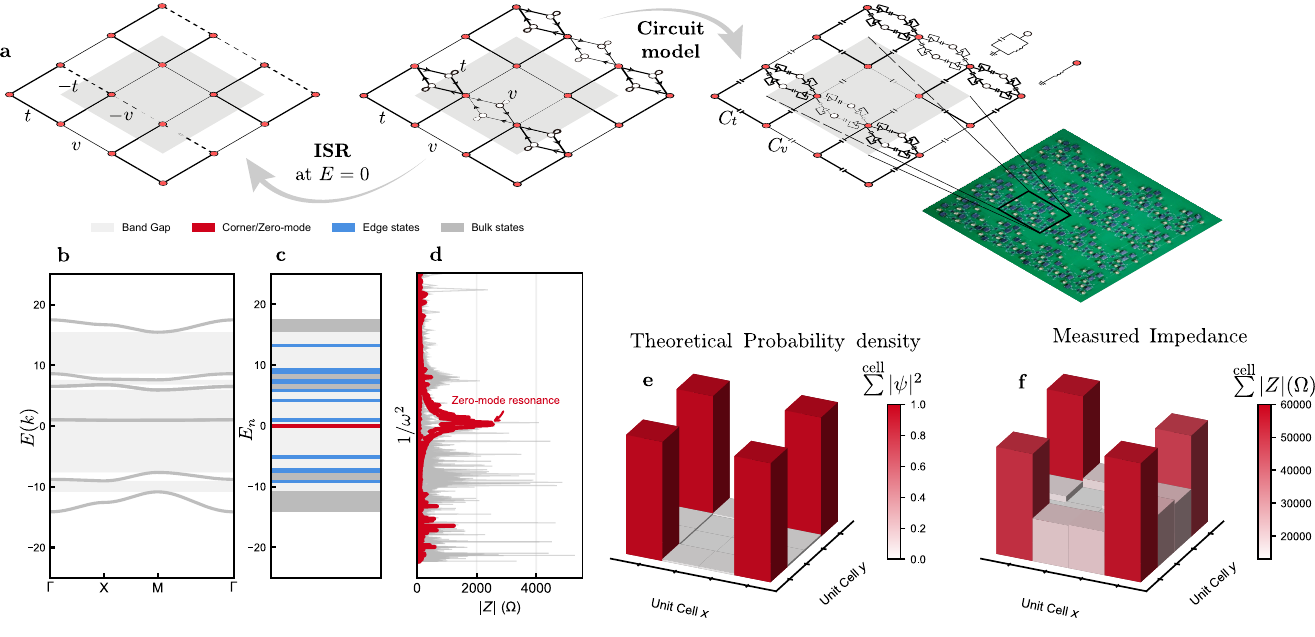}
    \caption{\textbf{Non-Hermitian Topological Circuit Realization and Characterization.} \textbf{a}, Schematic illustration of the non-Hermitian topological circuit realizing a two-dimensional corner state tight-binding model. The setup is similar to the Hermitian case, showing the tight-binding lattice model, its circuit mapping, and the physical circuit board implementation. \textbf{b}, Projected band structure ($E$ vs. $k$) of the non-Hermitian model with periodic boundary conditions (PBC). The parameters used are the intra-cell coupling $v=1$ and inter-cell coupling $t=8.2$. The gray areas denote the bulk states, and the white areas are the band gaps. \textbf{c}, Energy spectrum ($E_n$) of the $4 \times 4$ unit cell system (120 sites, $n=1, \dots, 120$) with open boundary conditions (OBC). The bulk states are in gray, edge states in blue, and the corner/zero-modes are highlighted in red. \textbf{d}, Experimental measurement of the inverse squared frequency ($1/\omega^2$) versus the measured impedance $|Z|$ ($\Omega$). The red peak marks the zero-mode resonance. \textbf{e}, Calculated unit-cell-resolved probability density $\rho(x_{\text{cell}}) = \sum_{i \in x_{\text{cell}}} |\Psi_i|^2$ for the four non-Hermitian topological states ($n=31$ to $34$). The summation over the 16 unit cells demonstrates the right-eigenstates localization at the corners. \textbf{f}, Measured spatial distribution of the frequency-integrated unit-cell impedance across the corner-state resonance band (centred at $f \approx 1.284$ MHz, spanning 1.198–1.3986 MHz), revealing robust localization consistent with non-Hermitian higher-order topological theory.
    }
    \label{non-Hermitian Figure}
\end{figure}
The latent non-Hermitian BBH model is constructed by adding sites 5 \& 6 with on-site hopping $v$ and sites 7 \& 8 with on-site coupling $t$ between sites 1 \& 4 from the original BBH model and making the hopping and coupling between them directed; cf. \cref{non-Hermitian Figure}~a. Consequently, the non-Hermitian system is governed by an $8 \times 8$ Bloch Hamiltonian $H$, where the non-Hermiticity is embedded through specific intra-cell hopping and inter-cell coupling configuration. The tight-binding Hamiltonian is expressed as
\begin{equation}
H =\left(\begin{array}{ c c c c c c c c}
0 & v+te^{-ik_{y}} & 0 & 0 & v & 0 & 0 & t \\
v+te^{ik_{y}} & 0 & v+te^{-ik_{x}} & 0 & 0 & 0 & 0 & 0\\
0 & v+te^{ik_{x}} & 0 & v+te^{ik_{y}} & 0 & 0 & 0 & 0\\
0 & 0 & v+te^{-ik_{y}} & 0 & 0 & v & te^{ik_{x}} & 0\\
0 & 0 & 0 & v & v & 0 & 0 & 0\\
v & 0 & 0 & 0 & 0 & v & 0 & 0\\
t & 0 & 0 & 0 & 0 & 0 & t & 0\\
0 & 0 & 0 & te^{-ik_{x}} & 0 & 0 & 0 & t
\end{array}\right) \,.
\end{equation}
By applying the ISR, we obtain the effective ISR Bloch Hamiltonian $H_{ISR}$, which maps the non-Hermitian model back to the original BBH model when $E=0$
\begin{equation}
H_{ISR} =\left(\begin{array}{ c c c c }
0 & v+te^{-ik_{y}} & 0 & -\frac{v^2}{v-E}-\frac{t^2}{t-E}e^{-ik_{x}} \\
v+te^{ik_{y}} & 0 & v+te^{-ik_{x}} & 0\\
0 & v+te^{ik_{x}} & 0 & v+te^{ik_{y}}\\
-\frac{v^2}{v-E}-\frac{t^2}{t-E}e^{ik_{x}} & 0 & v+te^{-ik_{y}} & 0 
\end{array}\right) \,.
\end{equation}

Interestingly, through the ISR, we obtain a Hermitian model from an underlying non-Hermitian one.
The resulting tight-binding model is shown in \cref{non-Hermitian Figure}a.
Figs.~\ref{non-Hermitian Figure}b,c display the projected energy spectra under periodic and open boundary conditions, respectively. While the periodic system exhibits a fully gapped bulk spectrum, the open system hosts additional in-gap states at $E=0$, corresponding to corner-localized modes of the finite lattice. The spatial structure of these states is illustrated in \cref{non-Hermitian Figure}e, which shows the unit-cell–summed probability density obtained from the tight-binding model at zero energy, revealing strong localization at the system corners.

We experimentally realized the model using an electric-circuit platform. The measured impedance spectrum is shown in \cref{non-Hermitian Figure}d, where a pronounced resonance associated with the corner modes appears near zero energy. The spatial structure of these modes is probed through the experimentally measured impedance profile at the resonance frequency, shown in \cref{non-Hermitian Figure}f. The observed impedance enhancement near the lattice corners closely matches the theoretically calculated probability density in \cref{non-Hermitian Figure}e, providing direct evidence for the emergence of higher-order topological corner modes in the system.

\subsection{Experimental realization}

The key ingredient in such non-Hermitian model is the unidirectional, non-reciprocal couplings. In the circuitry experiments, we could realize the non-reciprocal coupling via voltage followers by employing operational amplifiers (Texas Instruments LM6172IMX/NOPB).
The ideal voltage follower is made with an operational amplifier that has its output directly fed back to the inverting input, while the input signal is applied to the non-inverting input.
As a result, the output voltage closely follows the input voltage. However, the input is isolated from the output. In terms of circuitry Laplacian, it becomes evident for the exemplary circuit in the following, that the current on node b, $I_b$, relates both voltages $V_a$ and $V_b$ while $I_a$ is independent of $V_b$.
Thus, all components serially connected with the voltage follower become unidirectional, corresponding to the unidirectional couplings in the tight-binding model.

\begin{figure}
    \centering
    \includegraphics[width=0.5\linewidth]{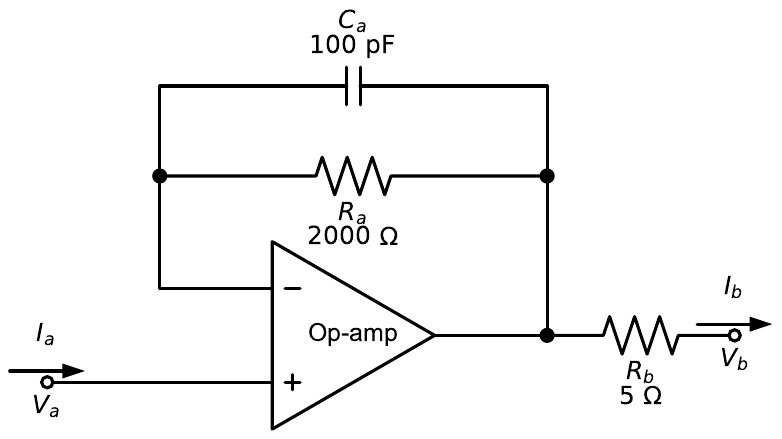}
    \caption{\textbf{Schematic of the voltage follower module used to implement undirectional coupling.} A parallel RC feedback network stabilizes the high frequency response, and a small series resistor $R_b$ at the output provides load isolation. Local power supply decoupling is applied.
    }
    \label{OpApm_figure}
\end{figure}

To achieve precise undirectional coupling in our experiments, we designed and fabricated a voltage follower module \cite{liuNonHermitianSkinEffect2021}, as illustrated in \cref{OpApm_figure}. In this circuit, the input signal directly drives the non-inverting input of the operational amplifier. To ensure high-frequency stability and to optimize the phase margin, a parallel feedback network consisting of a resistor $R_a = 2000~\Omega$ (0402WGF2001TCE) and a capacitor $C_a = 100~pF$ (Murata GRM0335C1E101FA01) is introduced between the output and the inverting input. Furthermore, an isolation resistor $R_b = 5~\Omega$ (0402WGF510KTCE) is connected in series at the output to physically isolate the load and prevent oscillations caused by capacitive loading. Regarding power management, a multi-state decoupling capacitor network is deployed on the printed circuit board to suppress high-frequency switching noise and power supply ripple.

Based on the unidirectional coupling module described above, we constructed a topolectric circuit representing the latent non-Hermitian BBH model. The tight-binding parameters including on-site energies, inter-cell coupling, and intra-cell hopping are systematically encoded in the admittance matrix $J$. To maintain consistency with the Hermitian system, we designed the on-site admittances for the retained latent BBH sites to be $J_{a,a} = i\omega(2t+2v) + 1/(i\omega L)$. To account for the auxiliary sites introduced by the non-Hermitian coupling, we define the intra-cell port admittance as $J_{b,b} = i\omega(2t) + 1/(i\omega L)$. Similarly, the inter-cell port admittances are set to $J_{c,c} = i\omega(2v) + 1/(i\omega L)$. In our experimental setup, the electrical components are uniformly chosen as $t=100$~pF (Murata GRM0335C1E101FA01), $v=820$~pF (Murata GRM0335C1E821FA01), and $L=10~\mu$H (CY105-100K). All experimental measurements were acquired using a vector network analyzer (Rohde \& Schwarz ZNL20), which allowed us to observe the robust spatial localization of the corner state impedance at the resonance frequency $f \approx 1.284$ MHz, providing direct evidence for the non-Hermitian higher-order topological protection in our circuit.

Similar to our realization of a partial latent Hermitian BBH model, we pruned the system at the upper and lower boundary, removing a total of $8$ sites for our system of $4\times{}4$ unit cells; the final system was then composed of $120$ sites.

In \cref{non-Hermitian Figure}d, \cref{non-Hermitian Figure}f, we show measurement results, which clearly show the presence of corner modes.

\section{Disorder}
\begin{figure}[b]
    \centering
    \includegraphics[width=\linewidth]{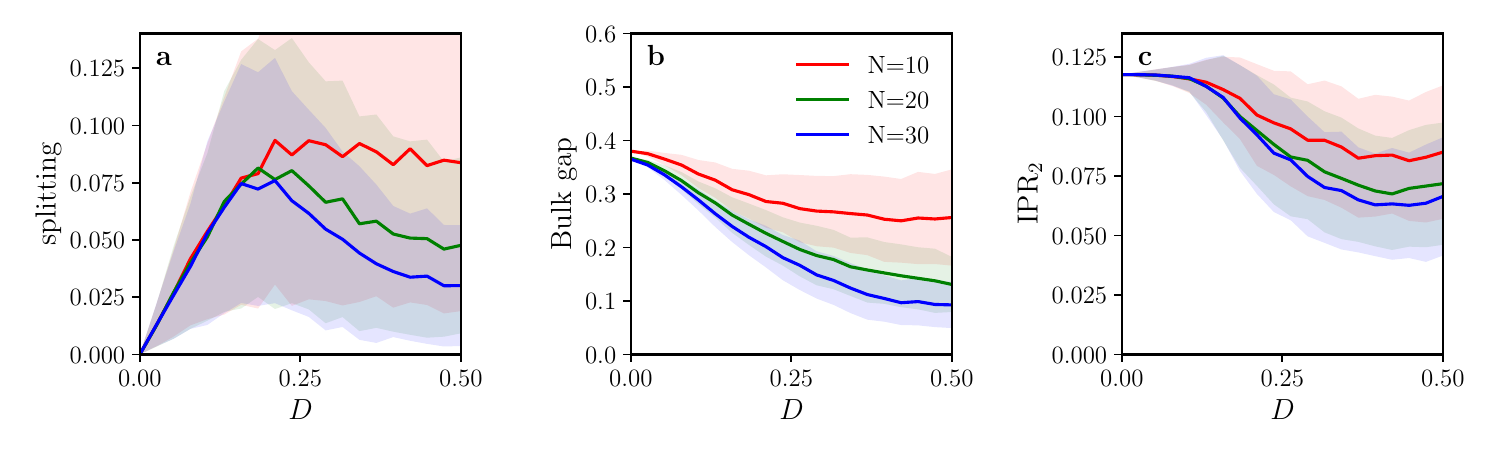}
    \caption{\textbf{Disorder averaged edge state splitting for symmetry-breaking disorder.} \textbf{a}, Edge-state-energy splitting, \textbf{b} bulk gap, and \textbf{c} $\text{IPR}_2$, for $R=500$ disorder realizations of hopping disorder in Eq.~\eqref{eq:PLSSH}. We take $t=3$, both $t$ and $D$ are measured in units of intracell hopping, which is set to $1$. The shaded background around the curves represents the standard deviation.}
    \label{fig:disorder}
\end{figure}
Finally, we investigate the role of disorder in systems protected by a partial latent symmetry. As an example, we consider the one-dimensional model introduced in Eq.~\eqref{eq:PLSSH}. We add hopping disorder by allowing the hoppings to fluctuate as
\begin{equation}
    t \rightarrow t+t_D,
\end{equation}
where $t_D$ is a random variable drawn from the uniform distribution $(-D,D)$. For each disorder strength, we compute the edge-state splitting, the bulk gap, and $\mathrm{IPR}_2$, averaging each quantity over $R$ disorder realizations.

The quantity $\mathrm{IPR}_2$ is a modified inverse participation ratio (IPR) that measures the localization of a pair of states. The conventional IPR is defined as
\begin{equation}
    \mathrm{IPR}(\psi)=\sum_i |\psi_i|^4,
\end{equation}
which equals $1$ for a state fully localized on a single site and approaches $0$ for delocalized states.

In the present case, however, there are two edge states, $\psi_L$ and $\psi_R$, with approximately degenerate energies. Any unitary combination of these states is therefore also an eigenbasis, i.e.
\begin{equation}
\begin{pmatrix}
\psi_1\\
\psi_2
\end{pmatrix}
=
U
\begin{pmatrix}
\psi_L\\
\psi_R
\end{pmatrix},
\end{equation}
with $U$ a $2\times2$ unitary matrix. Since in general
\begin{equation}
\mathrm{IPR}(\psi_L)+\mathrm{IPR}(\psi_R)
\neq
\mathrm{IPR}(\psi_1)+\mathrm{IPR}(\psi_2),
\end{equation}
the standard IPR depends on the chosen basis within the edge-state subspace. To avoid this ambiguity, we instead define the joint IPR
\begin{equation}
\mathrm{IPR}_2(\psi,\phi)=\frac{1}{4}\sum_i \left(|\psi_i|^2+|\phi_i|^2\right)^2,
\end{equation}
which is invariant under unitary rotations within the two-state subspace and therefore provides a well-defined measure of the localization of the edge-state pair.
Figure~\ref{fig:disorder}a shows the edge-state energy splitting as a function of disorder strength. Even for weak disorder, the two edge-state energies split apart. This occurs without a closing of the bulk gap, as shown in Fig.~\ref{fig:disorder}b, while the states remain relatively localized, see Fig.~\ref{fig:disorder}c. The origin of this splitting is that hopping disorder breaks the partial latent symmetry and thereby removes the associated topological protection. After performing the ISR, the hopping disorder is mapped onto on-site disorder in the effective SSH model.

\begin{figure}
    \centering
    \includegraphics[width=\linewidth]{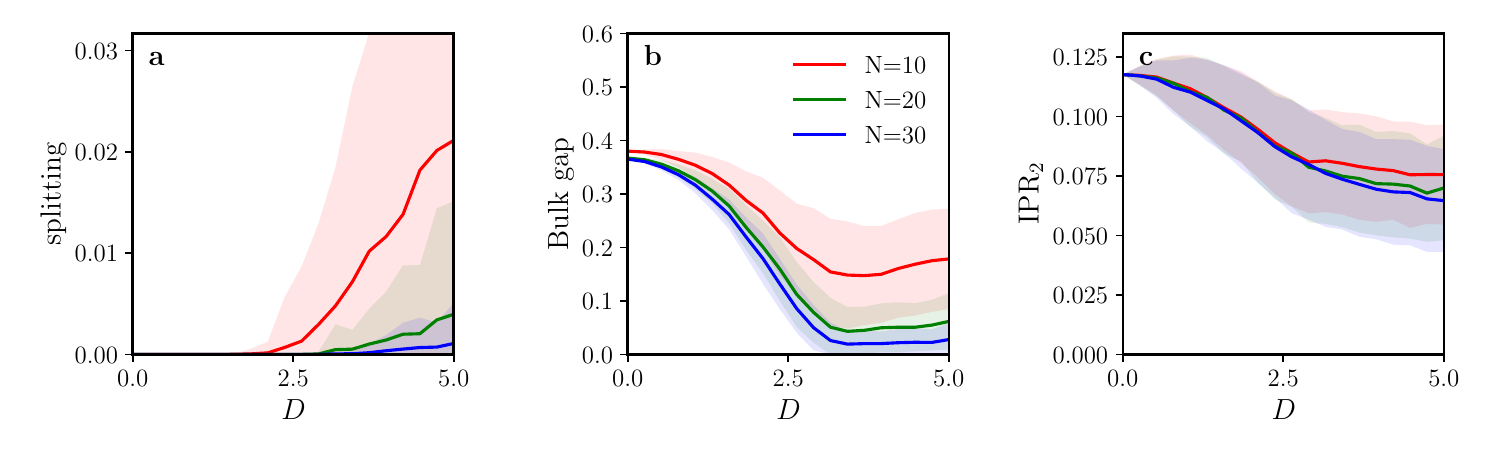}
    \caption{\textbf{Disorder averaged edge state splitting for symmetry-preserving disorder.}  \textbf{a} Edge-state-energy splitting, \textbf{b} bulk gap, and \textbf{c} $\text{IPR}_2$, for $R=500$ disorder realizations of partial latent symmetry respecting disorder in Eq.~\eqref{eq:PLSSH}. We take $t=3$, both $t$ and $D$ are measured in units of intracell hopping, which is set to $1$. The shaded background around the curves represents the standard deviation.}
    \label{fig:disorderrespect}
\end{figure}

The situation changes when considering perturbations that preserve the partial latent symmetry. For example, if we introduce perturbations connecting the $S$ sites in Fig.~\ref{fig:partialSSHBoundaries}, we obtain the behavior shown in Fig.~\ref{fig:disorderrespect}. In this case, the splitting remains small up to a critical value of $D$, beyond which the edge states separate in energy, see Fig.~\ref{fig:disorderrespect}a. This behavior is accompanied by a closing of the bulk gap, as shown in Fig.~\ref{fig:disorderrespect}b. The closing becomes exact only in the limit of large $N$. While the disorder does break the topological phase, the states remain localized, as can be seen in Fig.~\ref{fig:disorderrespect}c.

We conclude that topological insulators protected by partial latent symmetry behave similarly to conventional ones, in the sense that they remain robust against perturbations that respect the protecting (partial latent) symmetry.

\bibliographystyle{apsrev4-2}